\documentclass[sn-mathphys-num]{sn-jnl}
\usepackage{graphicx}%
\usepackage{multirow}%
\usepackage{amsmath,amssymb,amsfonts,amsthm}
\usepackage{mathrsfs}%
\usepackage[title]{appendix}%
\usepackage{xcolor}%
\usepackage{textcomp}%
\usepackage{manyfoot}%
\usepackage{booktabs}%
\usepackage{algorithm}%
\usepackage{algorithmicx}%
\usepackage{algpseudocode}%
\usepackage{listings}%

\newtheorem{theorem}{Theorem}
\newtheorem{remark}{Remark}%

\newtheorem{definition}{Definition}%
\newtheorem{lemma}{Lemma}

\usepackage{graphicx,tikz-cd}

\usepackage{dsfont}
\begin{document}

%\title[The mathematical framework of Euler Characteristic Surfaces: Correlation with Persistence Homology]{The mathematical framework of Euler Characteristic Surfaces: Correlation with Persistence Homology}


\author[1,2,3]{\fnm{Anamika} \sur{Roy}}

\author*[4]{\fnm{Atish} \sur{J. Mitra}}\email{atish.mitra@gmail.com}

\author[2,3]{\fnm{Tapati} \sur{Dutta}}

\affil*[1]{\orgdiv{Physics Department}, \orgname{Charuchandra College}, \orgaddress{ \city{Kolkata}, \postcode{700029}, \state{West Bengal}, \country{India}}}

\affil[2]{\orgdiv{Dept. of Physics}, \orgname{St. Xavier's College}, \orgaddress{\city{Kolkata}, \postcode{700016}, \state{West Bengal}, \country{India}}}

%affil[3]{\orgdiv{Condensed Matter Physics Research Centre}, \orgname{Jadavpur University}, \orgaddress{\city{Kolkata}, \postcode{700016}, \state{West Bengal}, \country{India}}}
\affil[4]{\orgdiv{Mathematical Sciences}, \orgname{ Montana Tech}, \orgaddress{\city{Butte}, \postcode{59701}, \state{Montana}, \country{United States}}}

\title[Exploring the Topology of  Dynamical Systems: Euler Characteristic Surfaces, Betti Surfaces and  Persistent Homology]{ Euler Characteristic Surfaces: A Stable Multiscale Topological Summary of Time Series Data}

\abstract{We present Euler Characteristic Surfaces  as a multiscale spatiotemporal topological summary of time series data  encapsulating the topology of the system at different time instants and length scales. Euler Characteristic Surfaces with an appropriate metric is used to quantify stability and locate critical changes in a dynamical system with respect to variations in a parameter, while being  substantially computationally cheaper than available alternate methods such as persistent homology. The stability of the construction is demonstrated by a quantitative comparison bound with persistent homology, and a quantitative stability bound under small changes in time is established.  The proposed construction is used to analyze two different kinds of simulated disordered flow situations. }


%\pacs[MSC Classification]{62R40, 55N31}

\maketitle
\section*{Declarations}

\begin{itemize}
%\item Funding {Not applicable}
\item \textbf{ Conflict of interest:} The authors have no competing interests to declare that are relevant to the content of this article.
%\item Ethics approval and consent to participate
%\item Consent for publication
\item \textbf{Data availability:} Data sets and codes generated during the current study are available from the corresponding author on reasonable request. 
\end{itemize}

\section{Introduction}
Studying data in terms of its topology has been a rapidly evolving field in recent years because of the advantages of understanding the shape of large data sets. Having initially started with explorations of static data clouds \cite{carlsson2009, adler2010, carlsson2007}, %{brain, minkowski, etc},  
 complex data from different kinds of dynamical systems are being studied with this approach, e.g. biological aggregation \cite{topaz2015}, social systems, gene networks \cite{mandal2020,rabadan2020}, economics/financial data \cite{gidea2018topological,yen2021using}, power engineering \cite{islambekov2020harnessing} etc. Often times when one lacks knowledge of the precise mechanisms that drive a large complex dynamical system, looking upon the physical image of the system as a time varying metric space in order to study the shape of its data points may give insight into the dynamics and parameters involved therein. In recent times in the field of Topological Data Analysis (TDA),  some work has been done on building topological tools to study the shape of complex dynamical data sets, e.g.,  crocker plots \cite{topaz2015, ulmer2019}, vineyards \cite{cohen2006vines} maps, crocker stacks \cite{xian2020}, zigzag persistence \cite{carlsson2009zigzag,myers2023} etc. Most of these tools can be computationally complex and stability results need to be further investigated.
 
This work elaborates on the idea of a  topological summary ``Euler Characteristic Surface" (ECS) introduced in \cite{roy2020}, that summarizes the shape information of a dynamical data set and thus helps in characterization of and distinction between dynamical systems \cite{roy2023}. We analytically establish its mathematical foundations and stability, and present applications in the form of  simulated dynamical models.

%\subsection{Topology and dynamics at multiple scales}

Data sets arising in the physical world are necessarily finite sets of points often embedded in a Euclidean space. As our purpose is exploring the shape of such data sets, we resort to a multi-scale approach to extract all possible meaningful information. Our approach is inspired by the philosophy inherent in multi-scale modeling of physical phenomena, and made rigorous by the mathematical ideas of topological persistence. 

\begin{figure}[h!]
\centering
\includegraphics[width=0.7\textwidth]{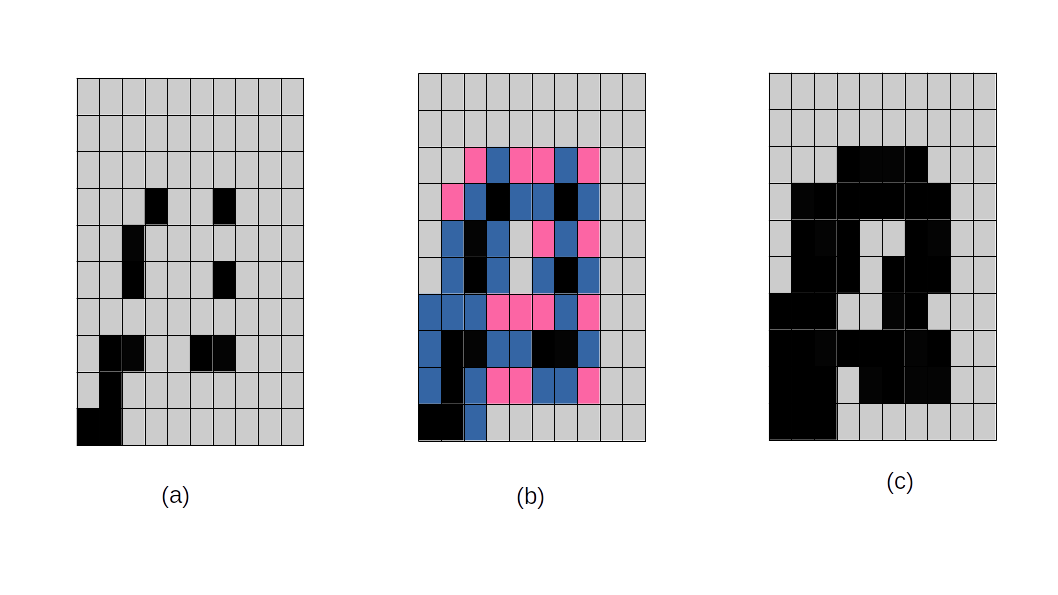}
\caption{Schematics of Coarse-graining/union of $r-neighborhood$ approach.(a)pixels at scale 0(no-coarse graining applied) (b)The nearest neighborhood marked with blue color and the next nearest neighbors marked with pink color, (c) coarse-graining applied at scale 1 with diagonal connections attached with 0.5 probability.}
\label{coarse-grain}
\end{figure}

Given a finite set of points $F=\{x_1,x_2, \cdots, x_n\} \subset \mathbb{R}^N$, we do a physically inspired ``coarse-graining" - where we replace a point by a ball of some non-trivial radius, the `scale'. More specifically, we  replace $F$ by the set $\displaystyle{F_r =\cup_{x \in F} B_r(x)}$, which is the union of closed balls of radius $r >0$ centered at each point of $F$. For pixelated images or discrete grid points, each black pixel/grid gets replaced by the union of that pixel with its nearest neighbors upto  its $r$-th neighborhood, Fig.(\ref{coarse-grain}). Unlike $F$, the set $F_r$ has interesting topology, and the topology changes with the value of $r$. If in addition, we assume the $x_i \in F$ to be time dependent, we can assume each $x_i$ to be a function of $t$, and our ``coarse-grained" set is represented as $F_{r,t}$.  Summarizing, our approach is to explore a spatio-temporal data set by studying the dynamic topology of $F_{r,t}$. The topological device we use to explore our data sets is based on the possibly simplest topological invariant - the Euler Characteristic.

The Euler Characteristic has been used in in multiple physical applications \cite{scholz2012,santos2019, dorso2012}, and the Euler Characteristic Curve  has been used for many years in studying the topology of pixelated digital images \cite{snidaro2003}. Our  past contribution in this area has been  encoding  both the physical scale of resolution and the time evolution of a dynamical phenomena as a single topological summary in its ECS. In the present work we continue our effort in quantifying the stability of our construction. A slightly different form of  Euler Characteristic Surfaces have been studied by different groups of authors \cite{beltramo2022, dlotko2023}, but those works do not consider time varying data.

Our technique is summarized in the following flowchart. Starting from a dynamical system $\{D\}$ represented by a dynamic point cloud $F_{t}$, we create the coarse graining cloud $F_{r,t}$,  from which create a family of complexes $\mathcal{K}_{r,t}$ which has the same homotopy as $F_{r,t}$. Then we create a topological signature of the dynamical system  by the Euler Characteristic Surface $\chi_{r,t}$.

\begin{equation*}\label{FromPointCloud2ECS}
\boxed{
    \mathcal{D}=F_{t} \longmapsto F_{r,t}=\cup_{x_t \in F} B_r(x_t) \longmapsto \mathcal{K}_{r,t} \longmapsto {\chi}(r,t)
}
\end{equation*}

  \subsection{Outline}
  
 In Section. \ref{framework} the detailed mathematical framework of our Euler Characteristic Surface (ECS) and the corresponding metrics - the Euler metric (EM) - is given. Geometrical simplicial complexes such as the Alpha Complex is defined and the viability of using it in the analysis of large data sets is justified in \ref{ecs}. Concepts of persistent homology and its metrics - the p-Wasserstein metric and the bottleneck metric  - that are often used to distinguish between data sets are discussed next in \ref{per},  followed by the pathway in establishing a relation between Euler Characteristic Surface and the persistence modules, and the stability of ECS under small perturbations of the data set proposed in \ref{PH}. We demonstrate the robustness of our proposed tools of ECS and EM on two simulated dynamical data sets - (i) with variable number of data points (ii) conserved number of data points in Section.\ref{real}. Subsection.\ref{ebf} (modified eggbeater flow model) compares the  ECSs constructed via coarse-graining and geometric simplicial complexes along with testing the stabilty of the Euler Metric as a measure to distinguish between and analyze dynamical data sets. Our second model (Vicsek model) is considered in  \ref{vis}, and there also we compare the ECS with  Euler metric with the above mentioned persistent homology tools. Thus this work establishes and validates the relations between Euler Characteristic Surface  and Persistent Homology, building the theoretical foundation of our tool.

\section{Mathematical Framework}\label{framework}

\subsection{Cell Complexes and Euler Characteristic Surfaces }\label{ecs}

A dynamical system here refers to the time evolution of a finite set of points in some domain in Euclidean space. Our proposed method studies the time evolution of the  topology of  such a set of points via the topological invariant Euler Characteristic $\chi$. To extract the multi-scale topology of the set of points which we consider the vertex set, we build cell complexes at  a scale $r \ge 0$. Cell complexes \cite{hatcher} are topological spaces built inductively by attaching higher dimensional cells to the ``lower dimensional skeleton'', where the ``0 dimensional skeleton'' is the vertex set i.e. the point cloud to be studied. A cell complex can be a simplicial complex which is commonly used for studying the topology of finite set of points. For the case of studying  pixellated digital images, the cell complex used can be the complexes built from the set of pixels using the physically inspired ``coarse-graining" or the ``union of $r$-neighborhood method" \cite{roy2020}. Simplicial complexes can be thought of either as \textit{geometric structures} in $\mathbb{R}^d$ where ''simplices'' (n dimensional versions of triangles, e.g. a point being $0-d$ simplex, a line being $1-d$ simplex, a triangle being $2-d$ simplex etc.) are assembled  following some specific rules, or as a purely combinatorial structure called an \textit{abstract simplicial complex}.

\begin{definition}\label{AbstractScx}
An abstract simplicial complex is a finite collection $K$ of finite non-empty sets such that if $\sigma$ is an element of $K$, so is every non-empty subset of $\sigma$. An element $\sigma$ of $K$ is called a simplex of dimension $|\sigma| -1$. The dimension of the entire simplicial complex $K$ is $\displaystyle{\max_{\sigma \in K} dim (\sigma)}$. A subset $L \subset K$ is said to be a sub-complex of $K$ if $L$ is also an abstract simplicial complex.
\end{definition}

A geometric simplicial complex is clearly an abstract simplicial complex, and any abstract simplicial complex can be viewed as a geometric simplicial complex (its geometric realization) \footnote{Every abstract simplicial complex of dimension $d$ can be geometrically realized in $\mathbb{R}^{2d+1}$}. Therefore, henceforth we will just identify and refer to them  as simplicial complexes. There are two commonly used methods of constructing simplicial complexes on a set of points, the Vietoris-Rips complex and the \v{C}ech complex. In our work we use a variation of the \v{C}ech complex construction, which is an example of a Nerve Complex.

\begin{definition}\label{Nerve}
For a finite set of subsets $U=\{U_1,U_2, \cdots, U_n\} $ in $\mathbb{R}^n$, the Nerve complex $\mathcal{N}(U)$ is defined by the rule, a subset $\sigma \subset U$ is a simplex iff $\cap _{i \in \sigma} U_i\neq \emptyset$.
\end{definition}

\begin{figure}[h!]
\centering
\includegraphics[scale=0.5]{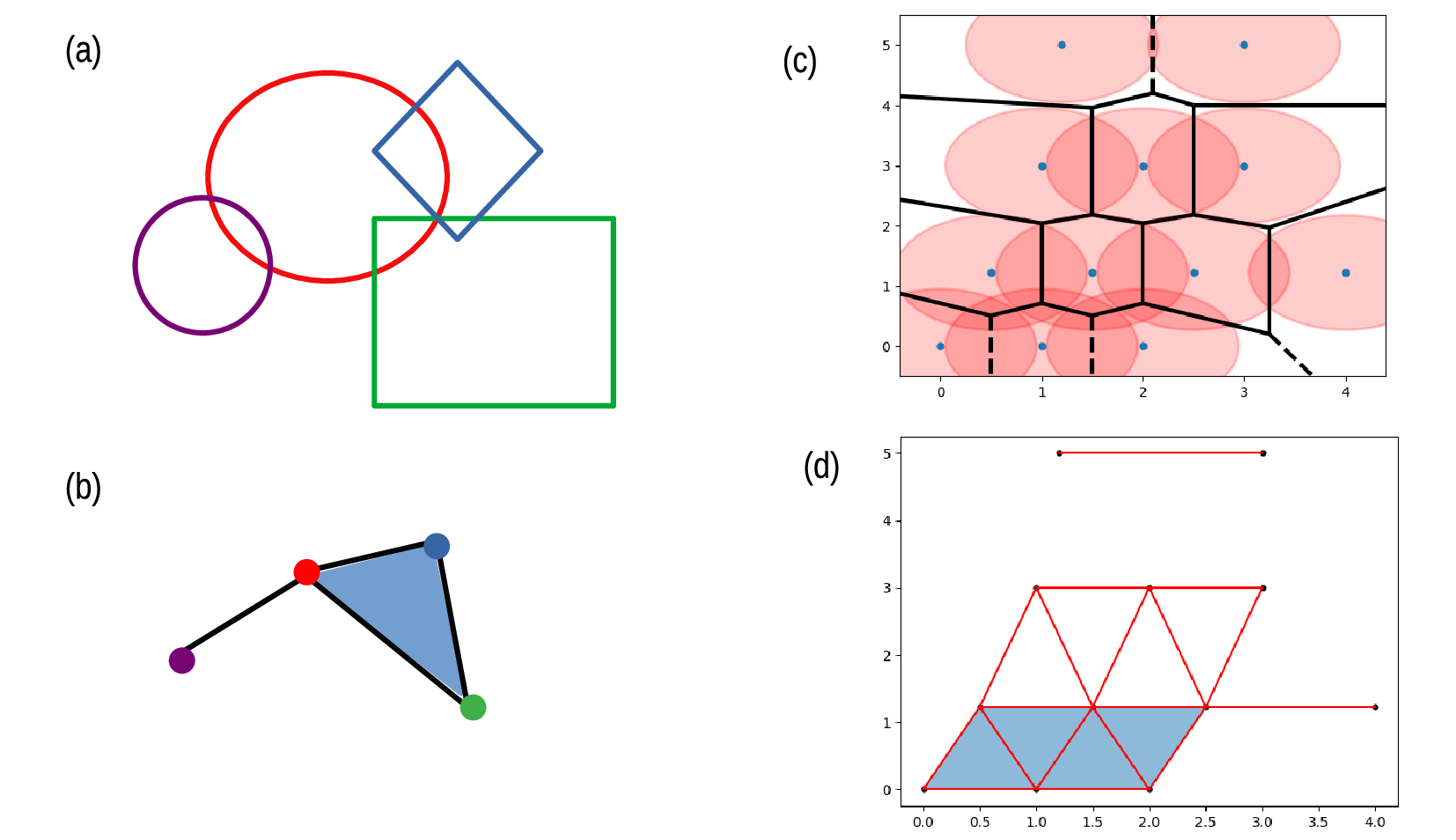}
\caption{ (a) The collection of subsets, (b) The corresponding Nerve complex, (c) Schematic to represent the intersection of Voronoi cells and balls of radius $r$ around a point cloud and (d) The corresponding Alpha complex.}
\label{complex}
\end{figure}

In Fig.(\ref{complex}a,b)the construction of Nerve complex for a given subsets is shown schematically considering the intersections of the subsets. Similarly, if we consider a set of points $P=\{x_1,x_2, \cdots, x_n\} \subset \mathbb{R}^d$, and $B(x_i,r)=\{x \in \mathbb{R}^d:  \|x_i-x\| \le r \}$ is the closed ball of radius $r$, then the \v{C}ech complex at scale $r$ is the simplicial complex $\text{\v{C}ech} (P,r) =\{\sigma \subset P: \cap_{x_i \in P} \ne \emptyset \}$. The \v{C}ech complex is the nerve of the closed $r$-balls around the set of points $P$ and  encodes the intersection pattern of the cells. 
Thus the \v{C}ech complex is homotopically equivalent to the $r$-neighborhood of $P$, the cell complexes constructed through coarse-graining.

The problem with working with the \v{C}ech complex is that, for a set of points in $\mathbb{R}^d$, the \v{C}ech complex constructed on it may have arbitrarily high dimension depending on the intersection pattern of the closed $r$-balls. This method can therefore lead to computational challenges. We shall resort to using the Alpha complex which bypasses this. For a finite set of points in $\mathbb{R}^2$, the Alpha complex can be defined as below \footnote{We  define the Alpha complex  in $\mathbb{R}^2$ as our applications are in that setting, but the definition works in $\mathbb{R}^n$}.
\begin{definition}\label{Alpha}
Consider a finite set of points in $F=\{x_1,x_2, \cdots, x_n\} \subset \mathbb{R}^2$, no four points lying on the same circle, and let $r>0$. The alpha complex at scale $r$ is defined as the nerve of $V_x \cap B(x,r)$, where $V_x$ is the Voronoi cell  corresponding to the point $x$. 
\end{definition}
Fig.(\ref{complex}c)shows the intersections between voronoi cells and balls of radius $r$ around a point cloud that generate the Alpha Complex shown in Fig.(\ref{complex}d). The alpha complex has the same homotopy type of $\displaystyle{F_r =\cup_{x \in F} B_r(x)}$, and its construction ensures that it is in $\mathbb{R}^2$ \cite{edels} \cite{virk2022}.\\
 \\
A dynamical system is a set of points in $\mathbb{R}^2$ at time $t$. In order to study the topological construct of the dynamical system we need to set up the simplicial complexes/cell complexes at scales $r \ge 0$ from the point clouds at different time instants. We examine these time-varying cell complexes through the lens of the Euler Characteristic. The Euler Characteristic of a cell complex can be defined as the alternating sum of the number of cells of each dimension.\\
If $\mathcal{K}$ is a finite dimensional cell complex, the Euler Characteristic is then defined by $\displaystyle{\chi(\mathcal{K}):= \sum_{i=0}^{\text{dim} (\mathcal{K})} (-1)^i \text{card} (\sigma_i)}$, where $\sigma_i$ is the collection of $i$-dimensional cells in $\mathcal{K}$. The above definition can be reformulated as an alternating sum of Betti numbers, from which it is evident that the Euler Characteristic is its homotopy invariant.

The idea of the Euler Characteristic Surface  and Euler Metric in the spatiotemporal setting was introduced while analyzing flow situations in drying droplets \cite{roy2020}, and was further explored in \cite{roy2023}. The Euler Characteristic Surface is defined as $\chi (\mathcal{K}_{r,t})$,  where $\mathcal{K}_{r,t}$ is the alpha complex at scale $r$ on the dynamical system at the time instant $t$. Here both parameters $r, t$ can vary continuously and we have a two dimensional surface as a map of the dynamical system.

 To quantify the difference between two ECSs describing two different dynamical systems, the $p$-Euler Metric  between the ECSs is defined  using the $L_p$ norm as :

\begin{equation}\label{Metric_eq}
\displaystyle{d(\chi_1,\chi_2) = \lVert \chi_1- \chi_2 \rVert_p= \Big(\int_{ [0,R]\times [0,T]}  |\chi_1 (r, t)- \chi_2 (r, t)|^p \Big)^{\frac{1}{p}}}
\end{equation}
with  $p=1,2$. With $p=2$, the Euler Metric (EM) is a  Hilbert space metric.
$\chi_{r,t}$ may be expressed as a  $m \times n$ dimensional matrix  with integer valued entries, where $m$ rows represent length scales and $n$ columns represent time instants.  For creating a binary image of the evolving data set, a grid mesh of appropriate size and shape  needs to be chosen to discretize the evolving system. A higher dimensional version of the Euler characteristic surface can be studied in a similar manner \footnote{In general, we can adapt  filtering functions  $f: \mathcal{K} \to \mathbb{R}^{N+1}$, with $N$ coordinates  representing $N$ spatial filters and one coordinate representing time}. Thus while the Euler Characteristic Surfaces works like topological signatures for dynamical systems, the estimation of the Euler Metric between two ECSs quantifies similarity or
dissimilarity between them and can act like a marker for critical transitions\cite{roy2023}.

\subsection{Homology, Betti Surfaces and Persistence Diagrams}\label{per}

Given a simplicial complex one can consider $k$-chains, a formal sum of $k$-simplices for a given dimension $d$, with coefficients in a convenient field of choice (we will consider $\mathbb{Z}_2$ coefficients). These $k$-chains equipped with the natural addition operation creates an abelian group $C_k$ of $k$-chains. We define the boundary of a $k$-simplex to be the sum of all its $(k-1)$-dimensional faces, and extend it to all $k$-chains. This gives a boundary homomorphism  $\partial_k$ from chain group $C_k$ to chain group $C_{k-1}$. We call a $k$-chain with empty boundary a $k$-cycle, thus creating a group $Z_k$ of $k$-cycles, which is a subgroup of $C_k$. On the other hand, we define a $k$-boundary to be a $k$-chain that is the boundary of a $(k+1)$-chain, getting a group $B_k$ of $k$-boundaries. From the fact that $\partial^2 = 0$ , we see that  that $B_k$ is a subgroup of $Z_k$. Finally, the $k$-th homology group $H_k$ is the quotient of the $k$-th cycle group modulo the $d$-th boundary group, and the $k$-th Betti number $\beta_k$ is the rank of this group. For further details on the theory of homology, see \cite{hatcher}.

In dynamical systems some features are observed to persist over several length scales indicative of some inherent significance.  In order to detect these special features, e.g. ``holes" or lack of matrix in an appropriate dimension in a simplicial complex arising from a point cloud, we can use the device of \textit{homology}; and to quantify the ``persistence'' of such holes, we use \textit{persistent homology}. In Topological Data Analysis (TDA) one of the most established tools is Persistent Homology. We give a very brief idea of this apparatus here. 

For a static point cloud and the associated Alpha complex construction, given a filtration(at different scales) of sub-complexes of $\emptyset \subset K_0 \subset K_1 \cdots \subset K_n=K$ we get a sequence of homology groups $ H_k(K_0) \to H_k(K_1) \to \cdots  \to H_k(K_n)$, and - for each $s<t$ - the induced homomorphism $f_k^{s,t}: H_k(K_s) \to H_k(K_t)$. The $k^{th}$ persistent homology group $H_k^{s,t}$ is the image 
of $f_k^{s,t}$, and the corresponding $k^{th}$ persistent Betti number $\beta_k^{s,t}$ is the rank of $H_k^{s,t}$. While the $k^{th}$ persistent Betti number quantifies how many of the homology classes (cycles) of $K_s$ still survives in $K_t$, the entire picture is graphically represented in a $k$-dimensional persistence barcode and $k$-dimensional persistence diagram. The $k$-dimensional persistence diagram is  a multi-set of points $(b,d)$ that records the $k$-dimensional homology classes that are `born' at filtration $B$ indicated by coordinate $b$ and `die' at filtration $D$ indicated  by coordinate $d$. For further details on the theory of persistent homology, see \cite{edels}.

For convenience of visualization, and for comparison with our Euler Characteristic Surfaces, we define - for an alpha complex at varying scale $r$  and fixed time $t_*$ - the $k$-dimensional Betti Curve $\beta_k(\mathcal{K}_{r,t_*})$. Similarly, we can define $k$-dimensional Betti Surfaces $\beta_k(\mathcal{K}_{r,t})$ and we have the relation $\displaystyle{\chi(\mathcal{K}_{r,t}):= \sum_{n=0}^{\text{dim} (\mathcal{K})} (-1)^n \beta_n(\mathcal{K}_{r,t}) }$

\begin{figure}[h!]
\centering
\includegraphics[width=1.1\textwidth]{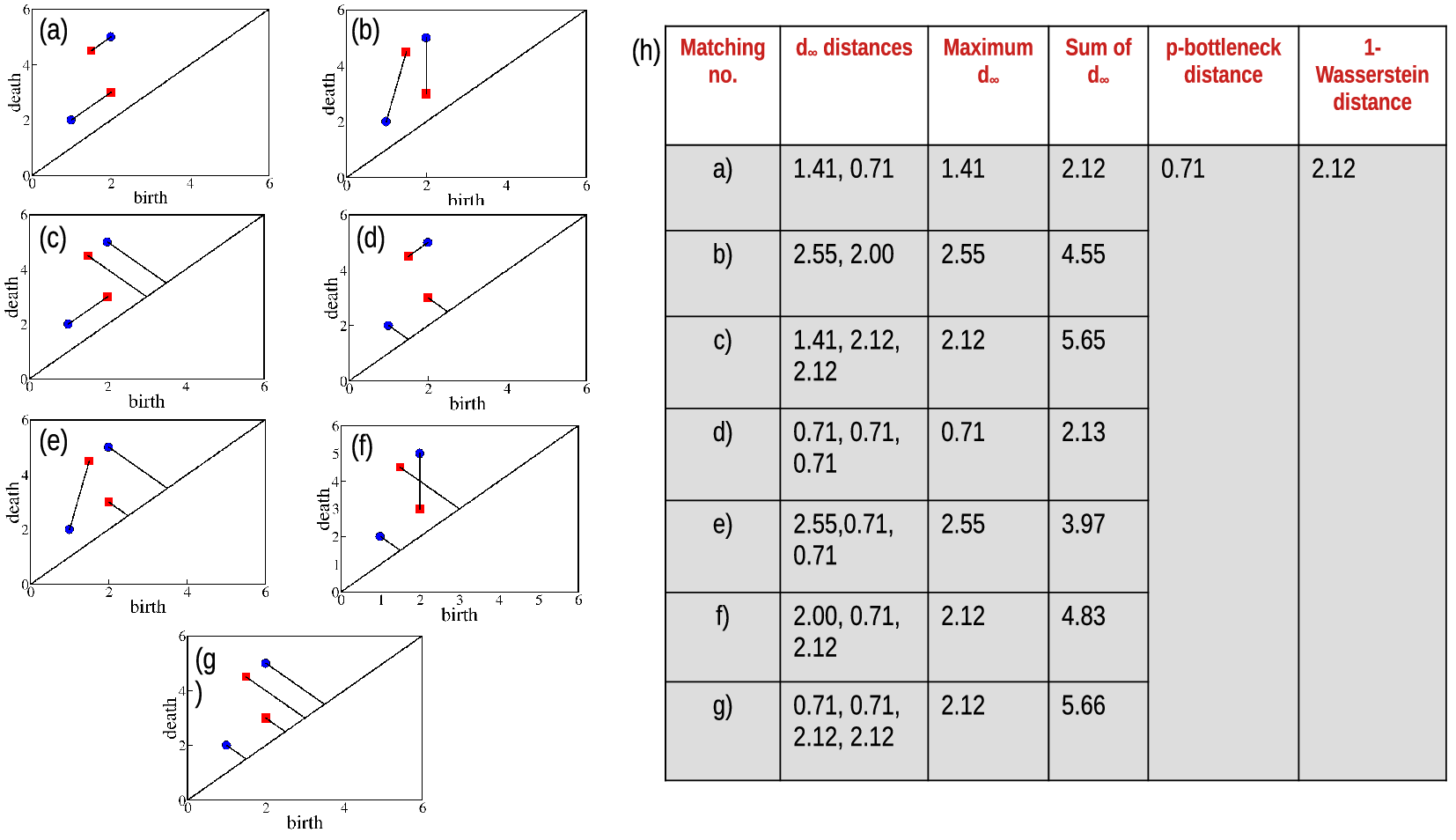}
\caption{ (a)-(g) represent different possible bijection or matching between two Persistence Diagrams(red and blue), (h) The table showcase the matching distances and the estimated metrics.}
\label{matching}
\end{figure}

Of the metrics that are usually considered on the space of persistence diagrams are the bottleneck metric and the $p$-Wasserstein metrics \cite{edels}. To estimate the difference, one creates bijections/matchings between two persistence diagrams (including diagonal points) and takes the infimum of all the matching distances in the appropriate norm ($\ell_{\infty}$ or $\ell_p$), Fig.\ref{matching}. In case of Bottleneck distance, the matching distance for each match is the maximum of the $d_{\infty}$ distances. For Wasserstein metric the $p$-th root of the sum of all $p$-th  powers of $d_{\infty}$ distances in a match is considered as the matching distance. Finally, in both cases the minimum of all the matching distances is considered as the metric value. The table in Fig.(\ref{matching}) portrays the step by step estimation of the above metric distances.
Formally, the $p$-Wasserstein distance ($p$ can be 1, 2, $\cdots$, $\infty$) between two $k$-dimensional persistence diagrams $\mathcal{A}$ and $\mathcal{B}$ is given as:  \footnote{Note that the ambient metric is the $d_{\infty}$  metric on the plane.}

\begin{equation}\label{p-Wasserstein}
W_p(\mathcal{A},\mathcal{B})= \inf_{\pi:\mathcal{A} \to \mathcal{B}}  \Big[ \sum_{(b,d) \in \mathcal{A}} \| (b,d) - \pi(b,d) \|^p_{\infty} \Big]^{\frac{1}{p}}
\end{equation}

\subsection{Comparison with Persistent Homology}\label{PH}

Though Persistent Homology is an well established topological tool in data science, it is computationally unwieldy and expensive in big data analysis, specially in the context of time series data that come from dynamical systems \cite{otter2017,kerber2016}, and the outputs of Persistent Homology are often not suitable for using directly into machine learning algorithms \cite{pun2022}. In this context our apparatus of SpatioTemporal Euler Characteristic Surface  (ECS) with the appropriate Euler Metric provides a computationally inexpensive tool. In this section we discuss the relation between our construction of Euler Characteristic Surface, Betti Surfaces  and Persistent Homology.

\subsubsection{Time-Slice wise stability of ECS against Persistent Homology}\label{TimeSliceStab}

In this section we discuss how techniques used in \cite{chung2022}  and  \cite{dlotko2023} gives the stability of a time-slice of the Euler Characteristic Surface (at a specific time $t=t*$) with the information obtained from the corresponding time-sliced Betti Surfaces and  hence from the corresponding persistence diagrams. This section is added here for completeness of the exposition, as our context is Spatiotemporal Euler Characteristic Surface (which deals with time series data, unlike in \cite{dlotko2023}), and to explain why the $L_1$ metric is the appropriate choice for ECS.

The main idea of the section is summarized in the following diagram:

\begin{equation*}\label{FromPD2EC}
\boxed{
\Big\{\textbf{PD}_n (\mathcal{K})\Big\}_{n=0}^{\infty} \longmapsto \Big\{\beta_n (\mathcal{K}_r) \Big\}_{n=0}^{\infty} \longmapsto \chi ({\mathcal{K}_r})
}
\end{equation*}
\vspace{.05 in}

Each of the two arrows above is associated with some loss of information. The sequence of Betti curves/surfaces can be  constructed from the corresponding persistence diagrams, and the Euler Characteristic curves/surfaces can be  constructed from the corresponding Betti curves/surfaces. Below we discuss how this loss of information occurs in a controlled way.

Given a simplicial complex $\mathcal{K}=\{\mathcal{K}_r\}_{r>0}$ with the filtration function determined by the scale $r$,\footnote{More precisely speaking, here we are considering the filtration function to be enumeration of the simplices in their order of appearance. See also Remark \ref{CriticalSimplices}.} let $\beta_n (\mathcal{K}_r)$ be the $n$-th Betti number of $\mathcal{K}_r$. Let $\textbf{PD}_n(\mathcal{K})$ be the $n$-dimensional persistence diagram of the filtered simplicial complex $\mathcal{K}$, and the complete persistent diagram is  $\displaystyle{\textbf{PD}(\mathcal{K})= \cup_{n=0}^{\textbf{dim} \mathcal{K}} \textbf{PD}_n(\mathcal{K})}$.

\begin{figure}[h!]
\centering
\includegraphics[scale=0.6]{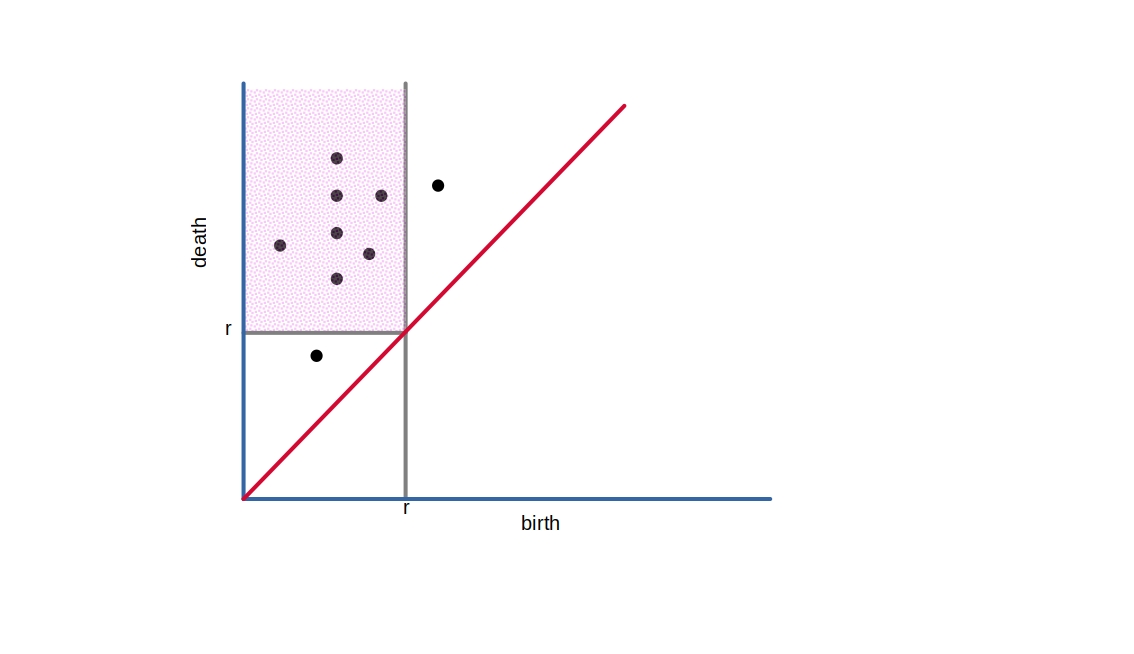}
\caption{a schematic of persistence diagram. The points within the shaded box whose birth at $\leq r$ and death at $> r$ contribute to the simplicial complex at scale $r$ and so to the Betti $r$ curve.}
\label{beta}
\end{figure}

As mentioned earlier, the Euler Characteristic of the simplicial complex $\mathcal{K}_r = \mathcal{K}(r, t*)$ with scale $r$  at time-slice $t*$ can be expressed as alternating sums of Betti numbers $\chi (\mathcal{K}_r) = \sum_n  {(-1)}^n \beta_n( \mathcal{K}_r)$.

\begin{definition}\label{BettiCurves-PD}

The $n$-dimensional Persistent Betti $r$-curve of $\mathcal{K}$ is defined as follows, where $\mathds{1}_{[b,d)} $ is the indicator function of the interval $[b,d)$.\footnote{ The indicator function is defined as,\\$\mathds{1}_{[b,d)}(x)=0 $ if $x \notin [b,d)$,\\$\mathds{1}_{[b,d)}(x)=1 $ if $x \in[b,d)$} 
\begin{equation*}
\beta_n (\textbf{PD}(\mathcal{K}), r) = \sum_{(b,d) \in \textbf{PD}_n(\mathcal{K})} \mathds{1}_{[b,d)} (r)
\end{equation*}

\end{definition}

\begin{lemma}\label{BettiCurvePD}
$\beta_n( \mathcal{K}_r) = \beta_n (\textbf{PD}(\mathcal{K}), r) $
\end{lemma}

\begin{proof}
From the Fundamental Lemma of Persistent Homology \cite{edels},  $\beta_n (\mathcal{K}_r) $ equals the number of points $(b,d)$ in the $n$-dimensional persistence diagram of $\mathcal{K}$ such that $ r \in [b,d)$ (see Fig.(\ref{beta})). 
\end{proof}

\begin{lemma}\label{IndicatorFunctionInequality}

For either of the configuration in figure \ref{indicator} (a) or (b), the following inequalities hold:

\begin{equation}\label{IndicatorforL1}
\lVert \mathds{1}_{[b^\mathcal{K},d^\mathcal{K})} (r) -  \mathds{1}_{[b^\mathcal{L},d^\mathcal{L})} (r) \rVert_1  \le 2 d_{\infty} ((b^\mathcal{K},d^\mathcal{K}),(b^\mathcal{L},d^\mathcal{L}))
\end{equation}

\begin{equation}\label{IndicatorforL2}
\lVert \mathds{1}_{[b^\mathcal{K},d^\mathcal{K})} (r) -  \mathds{1}_{[b^\mathcal{L},d^\mathcal{L})} (r) \rVert_2  \le \sqrt{2 d_{\infty} ((b^\mathcal{K},d^\mathcal{K}),(b^\mathcal{L},d^\mathcal{L}))}
\end{equation}

\end{lemma}

\begin{proof}
In each of cases \ref{indicator} (a) or (b), the inequalities \ref{IndicatorforL1} and \ref{IndicatorforL2} can be verified.  For example, for the configuration of figure \ref{indicator} (a), $\lVert \mathds{1}_{[b^\mathcal{K},d^\mathcal{K})} (r) -  \mathds{1}_{[b^\mathcal{L},d^\mathcal{L})} (r) \rVert_p^p = \int_{b^\mathcal{K}}^{b^\mathcal{L}} \lvert 1-0\rvert^p+ \int_{b^\mathcal{L}}^{d^\mathcal{K}} \lvert 1-1\rvert^p+ \int_{d^\mathcal{K}}^{d^\mathcal{L}} \lvert 0-1\rvert^p = \lvert b^\mathcal{L} - b^\mathcal{K}\rvert + \lvert d^\mathcal{L} - d^\mathcal{K}\rvert \le 2 d_{\infty} ((b^\mathcal{K},d^\mathcal{K}),(b^\mathcal{L},d^\mathcal{L}))$. A similar calculation works for for the configuration of figure \ref{indicator} (b).

\begin{figure}
\centering
\includegraphics[width =0.9\textwidth]{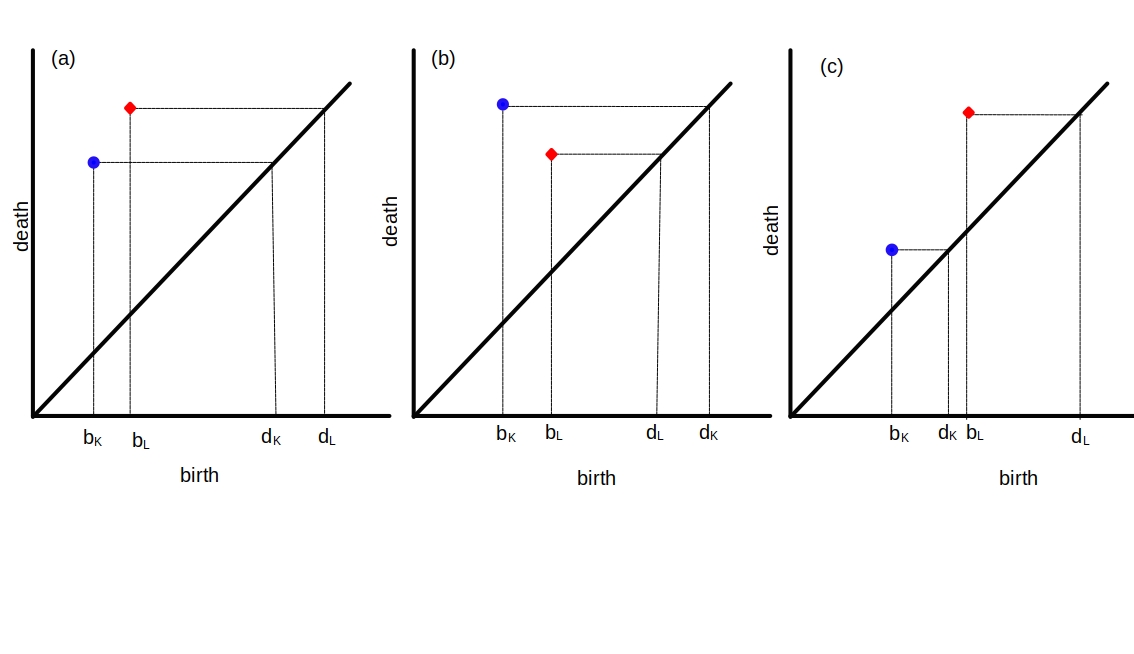}
\caption{(a) and (b) are the possible orientations for $(b_k,d_k)$ and $(b_l,d_l)$ for a possible  optimum matching between a pair of points from two different persistent diagrams. Matching (c) is not optimal.}
\label{indicator}
\end{figure}

\end{proof}

\begin{theorem} \label{SliceWiseStability-ECS}
Let $K$ and $L$ be simplicial complexes with non-zero persistence diagrams in a finite number of dimensions, and with each persistence diagram containing finitely many non-diagonal points. Then the Euler Characteristic Surface sliced at $t=t*$ with the Euler 1-metric and the Euler 2-metric satisfies the following inequalities  (the first one with respect to the 1-Wasserstein metric) at the finitely many non-zero persistence dimensions. In particular, where $N=\max\{\text{dim } \mathcal{K}, \text{dim } \mathcal{L}\}$,

\begin{equation}\label{SliceWiseStabilityEstimate-ECS-p=1} \footnote{See  \cite{dlotko2023} in the context of Euler  Characteristic Curves}
\| \chi(\mathcal{K}(r, t*)) - \chi(\mathcal{L}(r, t*)) \|_1 \le   2 \Big[\sum_{n=0}^N W_1(PD_{n} (\mathcal{K}),PD_{n} (\mathcal{L}))\Big]
\end{equation}

\begin{equation}\label{SliceWiseStabilityEstimate-ECS-p=2}
\| \chi(\mathcal{K}(r, t*)) - \chi(\mathcal{L}(r, t*)) \|_2 \le   \sum_{n=0} ^{N} \sum_i \sqrt{2 d_{\infty} ((b_i^\mathcal{K},d_i^\mathcal{K}),(b_i^\mathcal{L},d_i^\mathcal{L}))}\end{equation}

\end{theorem}

\begin{proof}

Following \cite{dlotko2023}, we consider an optimal matching between the two persistence diagrams with points $ \{(b_i^\mathcal{K},d_i^\mathcal{K})\},\{(b_i^\mathcal{L},d_i^\mathcal{L})\}$, with either the 1-Wasserstein or 2-Wasserstein metric, respectively.

For an optimal matching between two persistence diagrams with either the 1-Wasserstein or 2-Wasserstein metrics, configurations of pairwise matching between points in  figure \ref{indicator} (a) or (b) are possible. However, configuration \ref{indicator} (c) is not possible under neither of 1-Wasserstein or 2-Wasserstein metrics, as matching the points to the diagonal provides a better matching (as $|b_L-b_K|^p+|d_L-d_K|^p > |b_K-d_K|^p+|d_L-b_L|^p$ for $p=1,2$).

We have (using inequality \ref{IndicatorforL1}):

\begin{align*}
\| \chi(\mathcal{K}_r) - \chi(\mathcal{L}_r) \|_1  & = \lVert \sum_n  {(-1)}^n (\beta_n( \textbf{PD}(\mathcal{K}_r)) -  \beta_n( \textbf{PD}(\mathcal{L}_r)))\rVert_1  \\
& \le \sum_n \lVert  (\beta_n(\textbf{PD} (\mathcal{K}_r)) -  \beta_n( \textbf{PD}(\mathcal{L}_r)))\rVert_1  \\
& \le     \sum_{n=0} ^{N} \sum_i 2 d_{\infty} ((b_i^\mathcal{K},d_i^\mathcal{K}),(b_i^\mathcal{L},d_i^\mathcal{L})) \\
& \le   2 \sum_{n=0}^N W_1(PD_{n} (\mathcal{K}),PD_{n} (\mathcal{L}))
\end{align*}

We have (using inequality \ref{IndicatorforL2}):

\begin{align*}
\| \chi(\mathcal{K}_r) - \chi(\mathcal{L}_r) \|_2  & = \lVert \sum_n  {(-1)}^n (\beta_n( \textbf{PD}(\mathcal{K}_r)) -  \beta_n( \textbf{PD}(\mathcal{L}_r)))\rVert_2  \\
& \le \sum_n \lVert  (\beta_n(\textbf{PD} (\mathcal{K}_r)) -  \beta_n( \textbf{PD}(\mathcal{L}_r)))\rVert_2  \\
& \le     \sum_{n=0} ^{N} \sum_i \sqrt{2 d_{\infty} ((b_i^\mathcal{K},d_i^\mathcal{K}),(b_i^\mathcal{L},d_i^\mathcal{L}))} \\
%& \le    \sqrt{2}\sum_{n=0} ^{N} \max\{\#PD_{n}(\mathcal{K}),\#PD_{n}(\mathcal{L})\}\Big[d_\mathcal{B}(PD_{n} (\mathcal{K}),PD_{n} (\mathcal{L}))\Big]^{\frac{1}{2}}
\end{align*}

\end{proof}

\begin{remark}\label{W1betterthanW2}
We note from Theorem \ref{SliceWiseStability-ECS} that the Spatiotemporal Euler Characteristic Surface with the Euler 1-metric has better stability properties with respect to the  corresponding persistence diagrams with the 1-Wasserstein metric, as inequality \ref {SliceWiseStabilityEstimate-ECS-p=1} is independent of the number of points of the persistence diagram. In comparison, for the Euler 2-metric,  we have an  inequality \ref{SliceWiseStabilityEstimate-ECS-p=2} that depends on the number of points of the persistence diagrams. This pattern follows the existing stability results for other topological summaries \cite{carriere2017} \cite{chung2022}, where the distance between the summaries are bound from above by  the  1-Wasserstein distance between the corresponding persistence diagrams.  %This observation is used is section \ref{TempStab} to get a stability result for small perturbations of the data set with time.
\end{remark}

\begin{remark}\label{CriticalSimplices}
The Spatiotemporal Euler Characteristic Surface is defined using a continuous filtration in terms of the scale parameter. In practice, the ECS is calculated at points of  a discrete subset of scales $S$, so a proper choice of $S$ is important to ensure that the discrete subset of scales preserves all the information about changes in the original continuous filtration. In particular, one needs to consider all the critical scales $r_1 < r_2 < \cdots < r_m$, where the critical scales are those where at least one new simplex is created (as the underlying Delaunay complex is finite, there are only finitely many critical scales). This needs to be kept in mind while discretizing  the left side of the inequalities in Theorem \ref{SliceWiseStability-ECS}, to ensure that inequalities \ref{SliceWiseStabilityEstimate-ECS-p=1} and \ref{SliceWiseStabilityEstimate-ECS-p=2} are satisfied.
\end{remark}

\subsubsection{Stability of ECS against perturbations of the data set }\label{TempStab}

In this section we discuss temporal stability of the Euler Characteristic Surface construction.  For finite dimensional discrete dynamical systems where continuity properties can be assumed, we have temporal stability of the Euler Characteristic  by  using a Wasserstein Stability theorem as described below.

For this section we view our setting to be finite simplicial complexes with sub-level filtrations based on (simplex-wise) monotone filtering functions \footnote{A filtering function $f: \mathcal{K} \to \mathbb{R}$ is said to be monotone when $\tau \subset \sigma$ implies that $f(\tau) \le f(\sigma)$}. For example, the Rips complex on a finite data set gives rise to such a filtration, where the filtration function is the diameter of the simplex. For our case, the \v{C}ech (Alpha) complex on a finite data set gives rise to such a filtration, where the filtration function is the radius of the smallest enclosing ball of a simplex (of the Delaunay triangulation).

A recent powerful and useful result is the Cellular Wasserstein Stability Theorem below. 

\begin{theorem}[Skraba -Turner, \cite{skraba2023}]\label{SkrabaTurnerCelluarWassersteinStability}

For monotone functions $f,g :\mathcal{K} \to \mathbb{R}$ on a finite simplicial complex $\mathcal{K}$,  we have:

\begin{equation}\label{SkrabaCelluarIneq}
W_p(\textbf{PD}_n(f), \textbf{PD}_n(g)) \le \Big [\sum_{\textbf{dim }\sigma \in \{n, n+1\}} \lvert f(\sigma) -g(\sigma) \rvert^p \Big]^{1/p}
\end{equation}

\end{theorem}

The above, along with results from section \ref{TimeSliceStab} gives a  temporal stability result for Euler Characteristic Surfaces which we give as theorem \ref{TemporalStabilityECS}. Below we will say a discrete dynamical system $X$ is (temporally) uniformly continuous if for each $\epsilon >0 $ there is $\delta >0$ such that for each point $x_t \in X$ we have  $d(x_{t+\delta},x_t) < \epsilon$.

\begin{theorem}\label{TemporalStabilityECS}
Let $X_t=\{x_1(t), \cdots, x_M(t)\}$ be a (temporally) uniformly continuous discrete dynamical system with finitely many points in $\mathbb{R}^2$. Then for a sufficiently small  $0< \epsilon <  \frac{1}{2} \min \lVert x_i(t_0)-x_j(t_0)\rVert_2$, there is $\delta > 0 $  such that  if $\mathcal{K}(r,t)$ be the \v{C}ech filtration \footnote{Note that the \v{C}ech and Alpha filtrations give the same persistent homology, by the nerve thereom \cite{edels}.} on $X_t$, we have  $\| \chi(\mathcal{K}(r, t_0)) - \chi(\mathcal{K}(r, t_0+ \delta)) \|_1 \le \frac{1}{3} M(M+1)(M+2)\epsilon $. 
\end{theorem}

\begin{proof}

We note that for any acute angled  triangle $T$ in $\mathbb{R}^2$ and  for any $\epsilon > 0$, there is an $\delta_T >0$ such that after  perturbing the vertices by $\delta_T$, the circumradius of $T$ is changed by   at most $\epsilon$ (this follows from the continuity of the circumradius function of a non-degenerate triangle).  

Let $\epsilon > 0$ be less than  $ \frac{1}{2} \min_{1\le i \ne j \le M} \lVert x_i(t_0)-x_j(t_0)\rVert_2$.  Choose $\delta>0$ such that for $x_i(t) \in X_t$ we have  $d(x_i(t_0+\delta),x_i(t_0)) < \min{\delta_T}$, where the  minimum is over all acute angled triangles formed by points of $X_{t_0}$. Take the filtering function on a simplex  in theorem \ref{SkrabaTurnerCelluarWassersteinStability} to be the radius of the smallest enclosing circle of the points defining the simplex, with $f$ corresponding to points at time $t_0$ and $g$ corresponding to points at time $t_0+ \delta$.

Then, by theorem \ref{SkrabaTurnerCelluarWassersteinStability}, we have for $n=0,1$ the following inequalities (for the inequalities below, note that the smallest enclosing circle of a finite set of points in $\mathbb{R}^2$ is realized by either two of those points (diametrically opposite) on the circle, or by three of those points on the circle):

\begin{align*}
W_1(\textbf{PD}_0(f), \textbf{PD}_0(g))  \le {M \choose 1} \epsilon + {M \choose 2} \epsilon 
= {M+1 \choose 2} \epsilon = \frac{1}{2} M(M+1) \epsilon
\end{align*}

\begin{align*}
W_1(\textbf{PD}_1(f), \textbf{PD}_1(g))  \le {M \choose 2} \epsilon + {M \choose 3} \epsilon 
= {M+1 \choose 3} \epsilon = \frac{1}{6} M(M^2-1) \epsilon
\end{align*}

Finally, from theorem \ref{SliceWiseStability-ECS} we have:

\begin{align*}
\| \chi(\mathcal{K}(r, t_0)) - \chi(\mathcal{K}(r, t_0+ \delta)) \|_1   \le    2 \sum_{n=0}^1 W_1(PD_{n} (\mathcal{K}),PD_{n} (\mathcal{L})) = \frac{1}{3}M(M+1)(M+2) \epsilon
\end{align*}

In the inequality above we  only sum over $n=0,1$ as we know that the \v{C}ech filtration has the same persistent homology as the Alpha filtration.

 \end{proof}

\subsubsection{A note on Computational Advantages of ECS}

In view of the stability results initiated above, we note the definite computational advantages of the Euler Characteristic Surface with respect to persistent homology. While persistent homology has computational complexity of order about $\mathcal{O}(n^3)$ (where $n$ is the number of simplices), the Euler Characteristic Surface has computational complexity of order about $\mathcal{O}(n)$. This gives a  definite reason for exploring the possible applications of this construction in various practical situations involving big data.

\section{Application on simulated data}\label{real}
To establish the theoretical background explained in the previous section we use two different kind of time varying point sets generated from two kind of simulated systems. We illustrate on how the construction of ECS gets affected with perturbations and the choices of complexes in \ref{ebf} where the number of points in the vertex set grows with time. Followed by that we verify the relations established earlier in \ref{TimeSliceStab} and \ref{TempStab} with simulated data sets in \ref{vis} where the number of points in vertex set is constant in every time-step. 

\subsection{Eggbeater Flow: Dependence on Parameters, Stability of ECS }\label{ebf}
In a previous study \cite{roy2023} the authors had analysed simulated flow patterns of a fluid mixing model by constructing Euler Characteristic Surfaces using ``coarse-graining" or the ``union of $r$-neighbourhood"  approach. Here, we further extend the study on ECS and its mathematical characteristics using the fluid mixing model of modified egg beater flow \cite{franjione1992}. The simulation generates a point cloud at each time step following the equation Eq.(\ref{growth}) that describes the path of the flow at that time instant within a Poincare section of length $1\times1$.
\begin{equation}
\begin{aligned}
x_{t+1} = x_t -k y_t(1-y_t)\\
y_{t+1} = y_t -k x_t(1-x_t)
\end{aligned}
\label{growth}
\end{equation}

\begin{figure}[h!]
\centering
\includegraphics[width=0.75\textwidth]{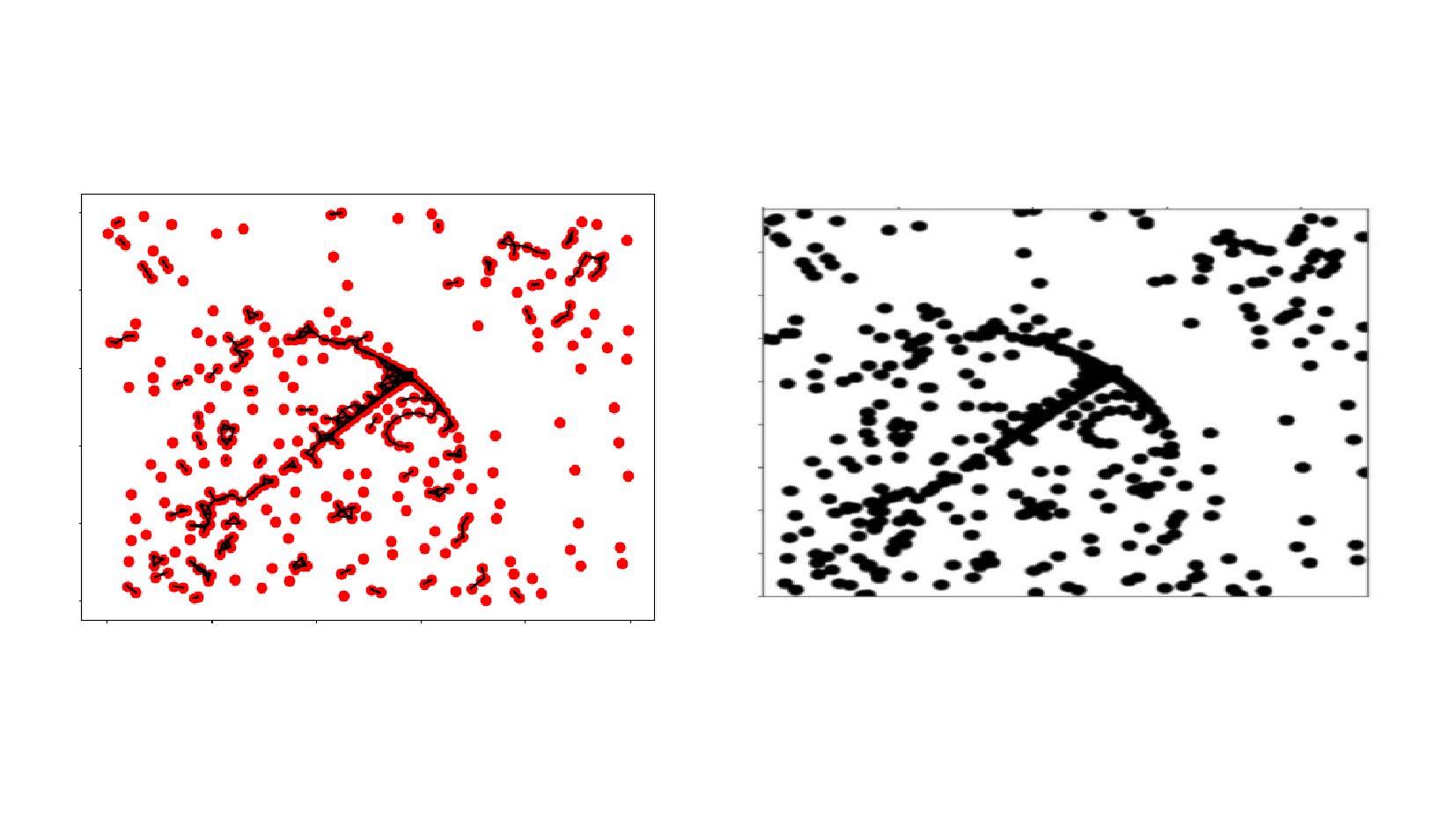}
\caption{(a) Alpha Complex of flow pattern of modified egg-beater flow at time-step t=500 with k=4.1,at scale r=$0.01447$ estimated $\chi$=161, (b)Same point set at same scale($r=10$) with ``coarse-graining" or ``union of $r$-neighbourhood" approach, estimated $\chi$ =148}
\label{chi_alpha_cluster}
\end{figure}

The seed point or starting  point is (0.9, 0.2) and the process is allowed until 10,000 time steps for a constant value of $k$. We produce a family of dynamical systems for different values of $k$ ranging between $k=4.0$ to $k=5.0$. This choice of parameter $k$ was done since for smaller value of $k<4$, the flow pattern is almost linear and not well spread over the domain while larger value of $k$ give almost uniform aggregation spread over the domain lacking diversity of topological features (further details are explained in \cite{roy2023}). The point cloud at each time step of the simulated flow system for a constant $k$ is used to make the Alpha complex at scales that are almost comparable to the scales used for discrete grid systems where we had built the cell complex via ``union of r-neighbourhood" or ``coarse-graining" approach. The integer values of the scales indicating the ``$r$-neighborhood" are converted into Euclidean distances for scales in Alpha Complex by considering the approximate pixel/grid size. The Euler Characteristic($\chi$) is estimated in Alpha complexes using alternative sum of numbers of zero dimensional, one dimensional, and two dimensional simplices. In our case, the complexes have two homology groups, $H_0$ and $H_1$. In case of ``coarse graining" method the value of Euler Characteristic was estimated by the difference between the number of black clusters($N_b$) and the number of white clusters ($N_w$) in the binary pixelated cell complexes, $\displaystyle{\chi= N_b - N_w}$. Fig.(\ref{chi_alpha_cluster}) shows the estimated values of Euler Characteristic ($\chi$) in both approaches. The values show a small deviation at higher scales (poorer resolution) as the conversion of pixel size to Euclidean distance is not precise at higher scales. For Alpha complexes, the estimation of Euler characteristic becomes exact at all scales since there is no discretization required.  

\begin{figure}[h!]
\centering
\includegraphics[width=0.99\textwidth]{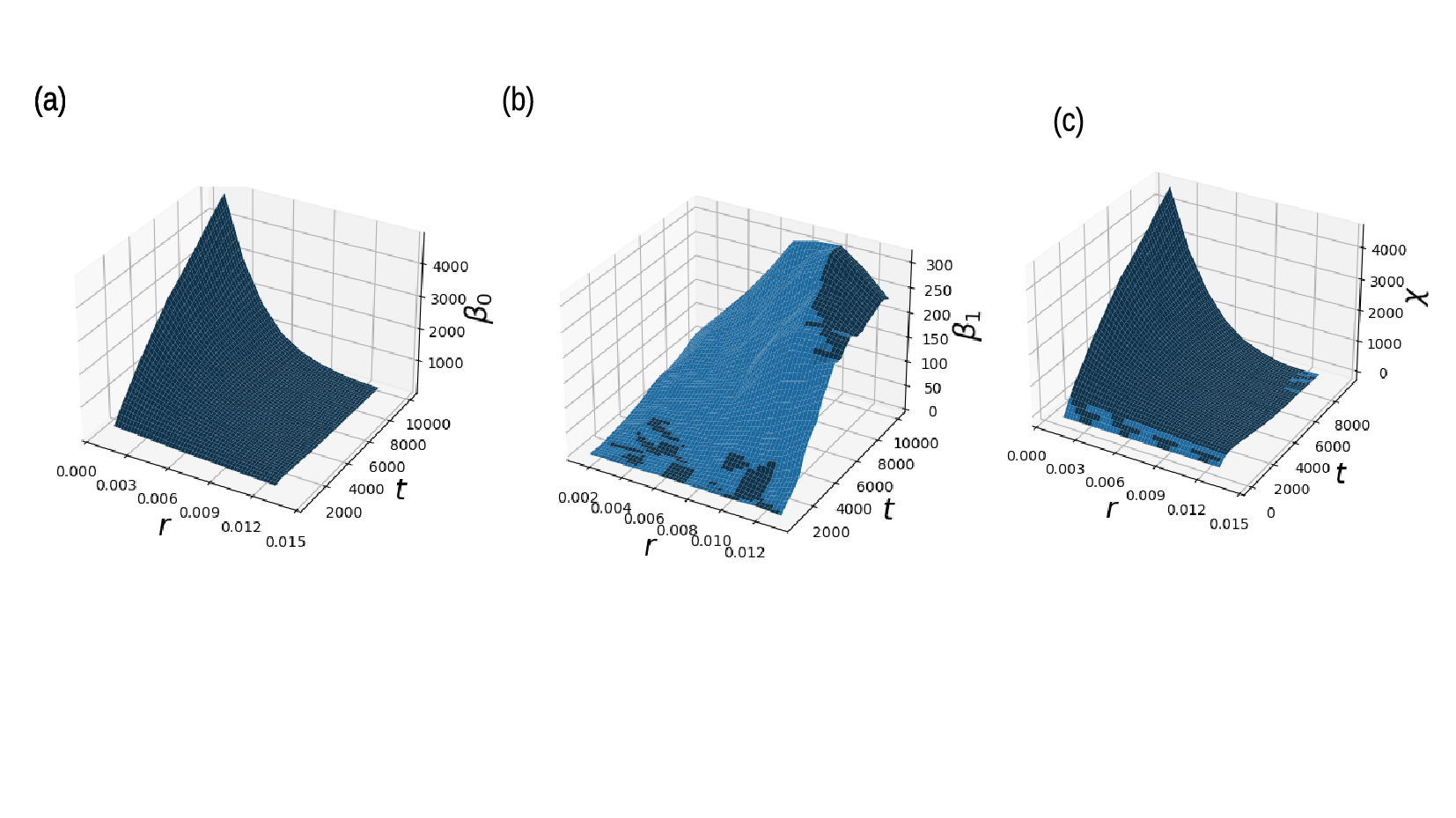}
\caption{ For Flow pattern with $k=4.1$ -(a) $0$-dimensional Betti surface, (b)$1$-dimensional Betti surface, (c) The Euler Characteristic surface. }
\label{betti_chi}
\end{figure}

\begin{figure}[h!]
\centering
\includegraphics[width=1.1\textwidth]{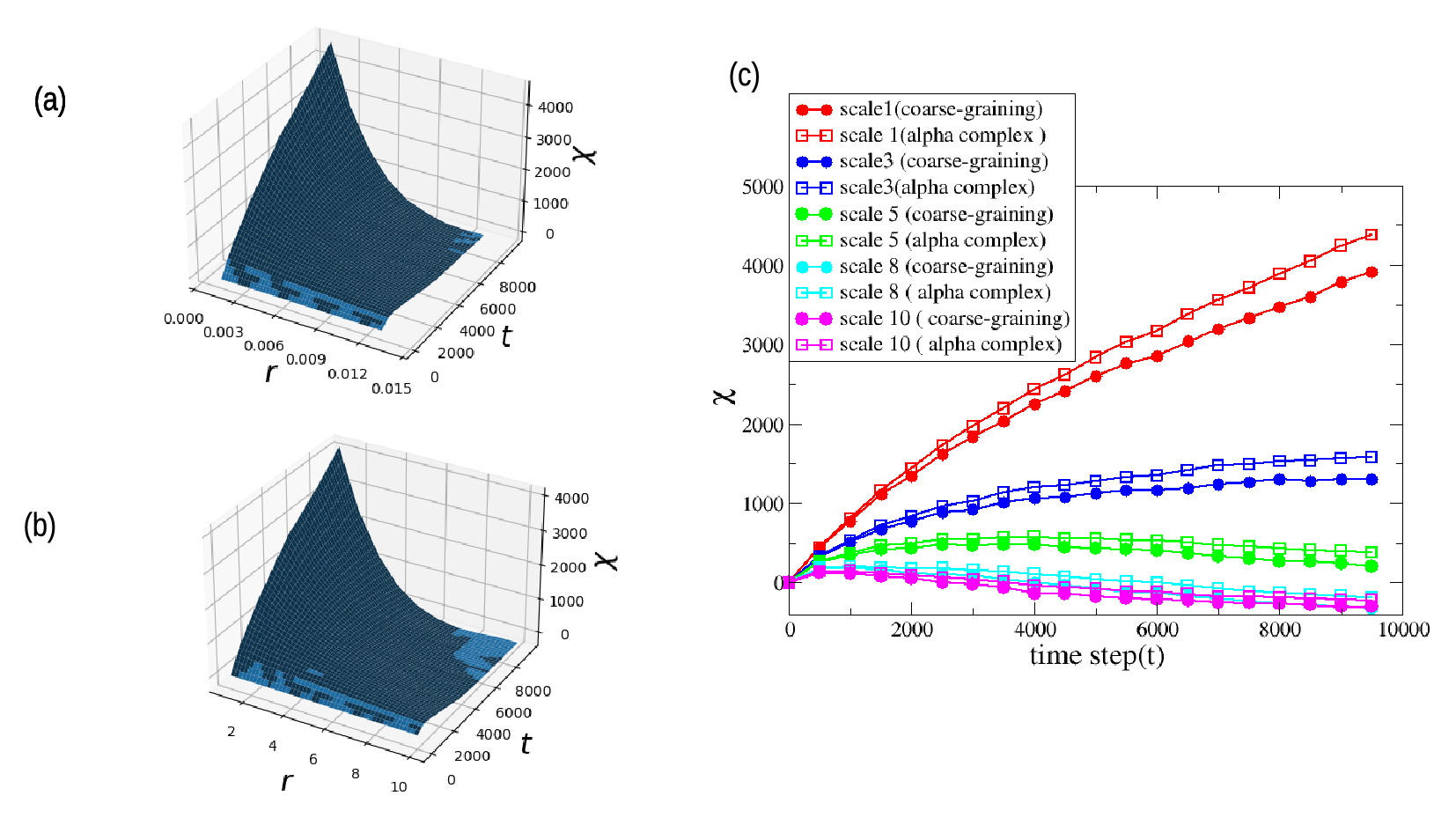}
\caption{ (a)Euler characteristic surface constructed through Alpha complexes,
(b)Euler characteristic surface constructed through 'coarse-graining' or union of r-neighborhood  approach on pixelated grid, (c) The values of estimated Euler Characteristic ($\chi$) at different scales using the two methods.}
\label{two_ecs}
\end{figure}
The Euler Characteristic Surfaces carry a summary of topological features present in all dimensions from the simplicial complexes  generated from a dynamical system. It must be noted that if one is interested to study a single dimensional topological feature in detail, for example the the 0-dimensional simplices representing the connected components/ clusters, the 1-dimensional simplices representing the loops, it can be done by constructing individual Betti surfaces similar to ECS construction. As seen in Fig.(\ref{betti_chi}), the spatiotemporal evolution of $n$-dimensional   Betti numbers can be summarised into surfaces that we name the $n$-Betti surface $\beta_n(r,t)$. For the simulated flow pattern discussed above, the Betti surfaces corresponding to critical point $k=4.1$ is displayed in Fig.(\ref{betti_chi}), where we can see the $\beta_0$ surface is qualitatively quite similar to the Euler Characteristic Surface but the $\beta_{1}$ surface representing the loops is clearly different and smaller in magnitudes. The metric distance between two Betti surfaces of two different dynamical systems can also be estimated in similar fashion that we do for Euler metric, the $L_2$ norm and the $L_1$ norm distance between two Betti surface of similar dimension.

\begin{figure}[h!]
\centering
\includegraphics[width=0.65\textwidth]{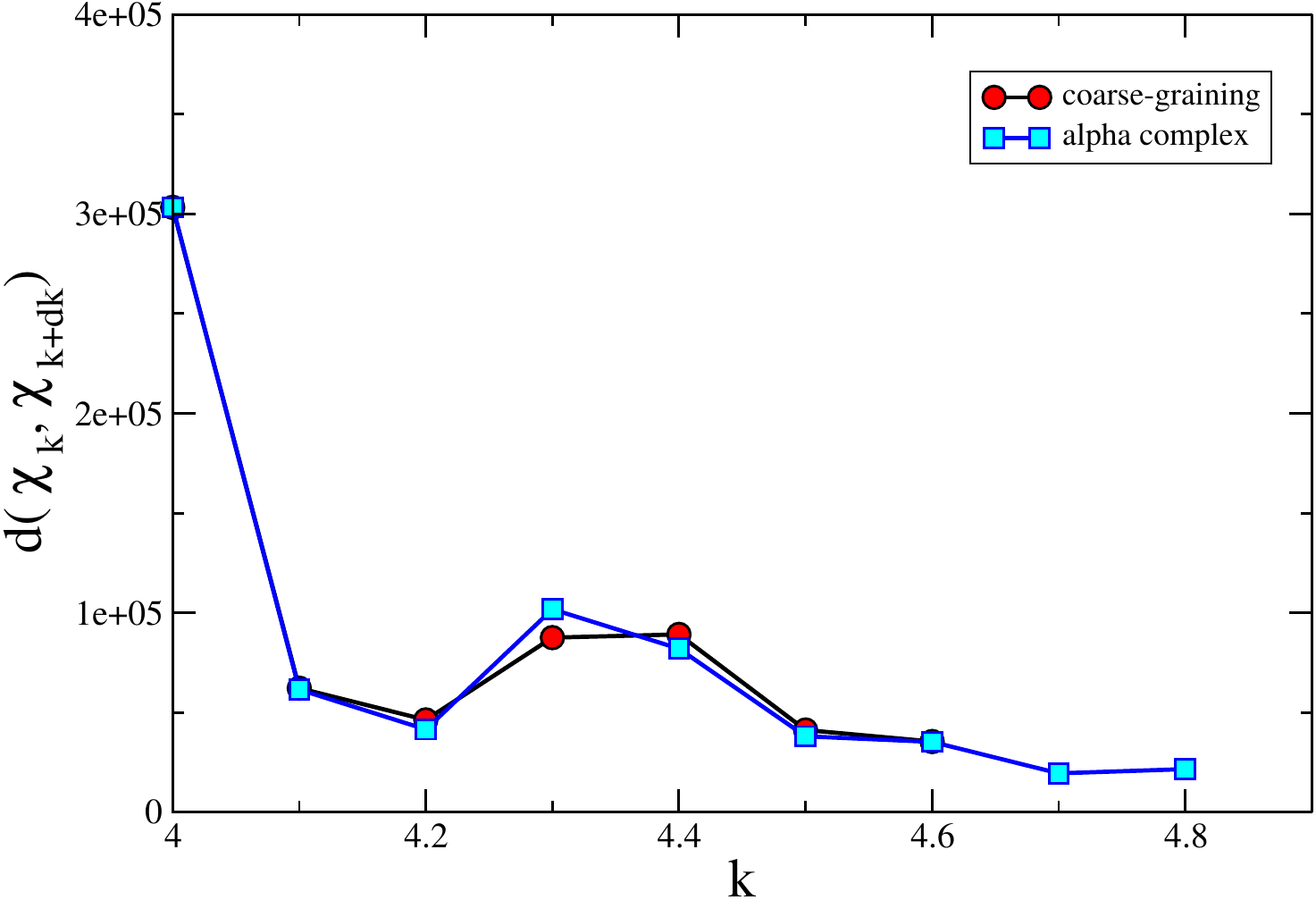}
\caption{ Euler metric estimated between surfaces with k and k+dk where dk=0.1, using both approaches.}
\label{em_alpha_cluster}
\end{figure}
On the overall, the Euler Characteristic Surfaces constructed from the two methods do seem to be similar, Fig.(\ref{two_ecs}). Further, their similarity or dissimilarity is quantified using Euler $L_p$ ($p=1,2$) Metrics. Euler Characteristic Surfaces are constructed for the dynamical systems of modified egg beater flow models using equations \ref{growth} with different $k$ values, using 'coarse graining' approach and the traditional Alpha Complex approach. We then estimate the Euler Metric between surfaces of dynamical systems corresponding to parameter $k$ and $k+dk$, for both cases. The results are plotted in Fig.(\ref{em_alpha_cluster}).The estimation of Euler Metric in both cases turns out to match with each other and the perturbation or error in estimation of Euler Characteristic matches with the corresponding Euler Metric. \textit{Thus the proposed Euler Metric is independent of the methods or approaches used to compute Euler Characteristic($\chi$). }

\begin{figure}[h!]
\centering
\includegraphics[width=0.65\textwidth]{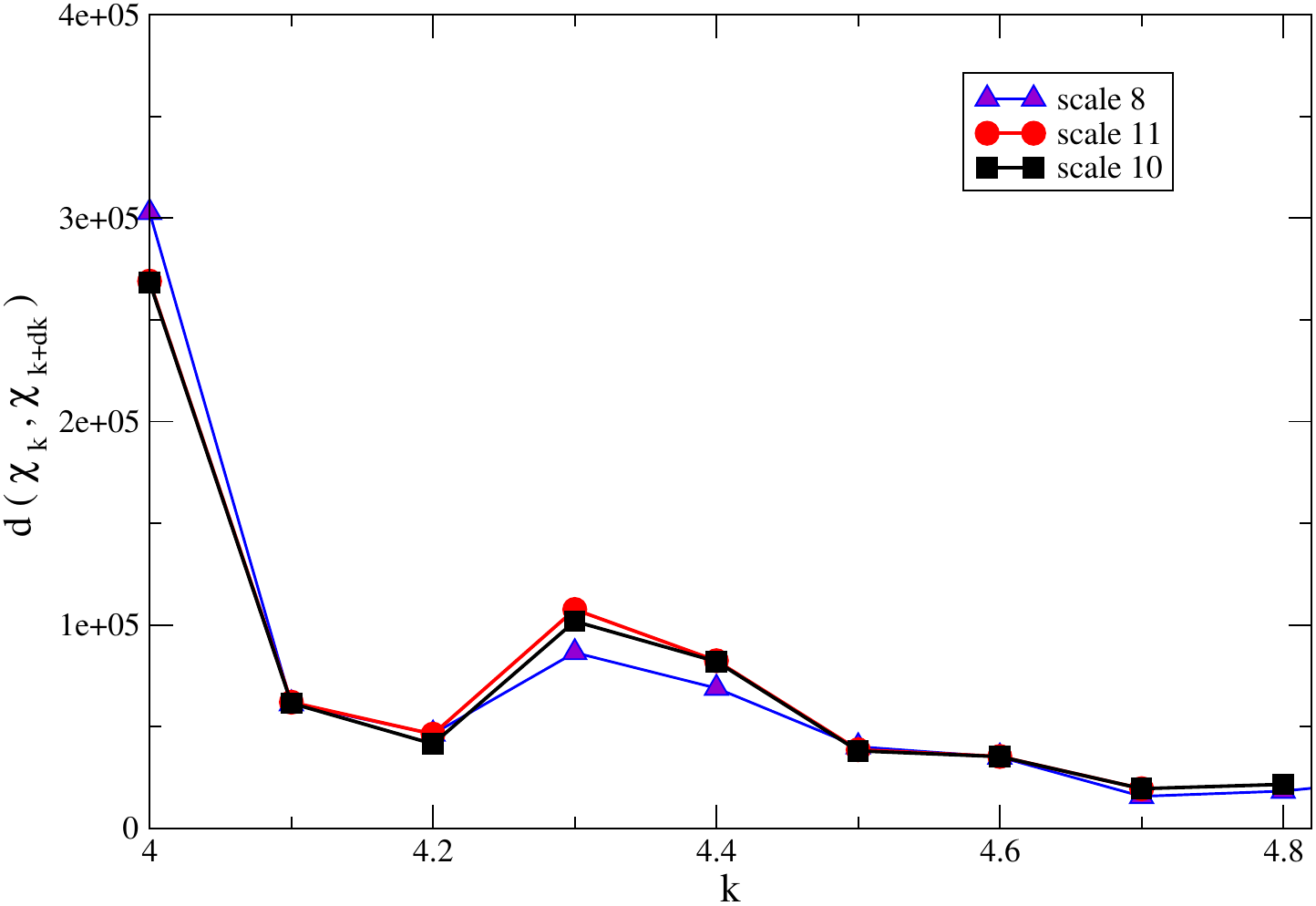}
\caption{ Euler metric estimated between surfaces keeping R upto scale 10(black)and upto scale 11 (red).}
\label{sc11_10}
\end{figure}

The stability of the Euler Characteristic Surfaces with respect to small perturbation in the scale of resolution is tested next. For a stable tool, small perturbation in the scale should result in small change in outcome or measure too. The question that comes to one's mind is how high the scale of filtration should be taken to construct the ECSs, and whether these surfaces are stable with respect to small perturbations. We therefore compare the Euler Metric estimated between two surfaces, one expanded upto scale $R$ with the other ECS expanded upto $R+dR$, keeping the other parameters constant. Given the change in scale $dR$ is small, the change in the measure of the Euler Metric turns out to be small too if the Euler Characteristic Surfaces are constructed with finer variations in scale and upto scales high enough to cover almost all possible topological features (in particular, critical scales should not be missed. See remark \ref{CriticalSimplices}). The results of perturbing the scale on Euler Characteristic Surfaces and thus on Euler Metric are displayed in Fig. (\ref{sc11_10}).

\subsection{Vicsek Model: A case study }\label{vis}
Different metrics  have been introduced in Persistent Homology to study the similarity and dissimilarity of shapes between two point sets, the popular ones being ``Bottleneck distance" and ``Wasserstein distance". In recent times there have been substantial works applying these tools on time varying point sets \cite{topaz2015}. In this section we test the stability of our proposed metric EM against one of these metric distances. For this, we study the alpha complexes of time varying point sets generated using \textit{Vicsek model}, widely studied to model collective dynamics of living matter \cite{vicsek1995}. In the complexes studied in section \ref{ebf}, number of points was not constant with time but in these point sets generated from Vicsek model, the number of points (living matter) in each time-step is kept constant. The classical Vicsek equation used to study collective motion is described in Eq.\ref{vis_eq} where each point/living matter travels with a constant magnitude of velocity($v$) and its orientation of velocity($\theta$) is influenced by its neighbouring particles along with a noise parameter $\eta$.
%\begin{equation}
\begin{subequations}
\begin{align}
\ x_i(t+1)=x_i(t) +v_i(t + \nabla t)\nabla t \\
\ \theta_i(t+\nabla t)= \frac{1}{N}\sum_{ \mid{x_i-x_j}\mid \leq R} \theta_j(t)+ U(\frac{-\eta}{2},\frac{\eta}{2})
\end{align}
\label{vis_eq}
\end{subequations}

%\end{equation}

We vary the distribution of the points in this model - (i) by varying noise parameter $\eta$ for a constant sphere of interaction with radius $R$ and (ii) by varying the radius of sphere of interaction $R$ for different noises ranging from $\eta =0$ to $5$. The box length $L$ was kept at $1.0$ unit, with number of particle $N=500$ moving with velocity of magnitude $\mid v \mid =0.03$. For lower value of noiseor zero noise the randomly distributed point set slowly transforms into aggregated clusters with order taking place in the system. When the noise is high, it dominates the neighborhood interaction and therefore chaotic regime exists in the point sets with no ordering observed. This is discussed in details in the following analyses.

\subsubsection{Analysing point cloud with and without noise($\eta$) for a constant $R$: }
\begin{figure}[h!]
\centering
\includegraphics[width=0.99\textwidth]{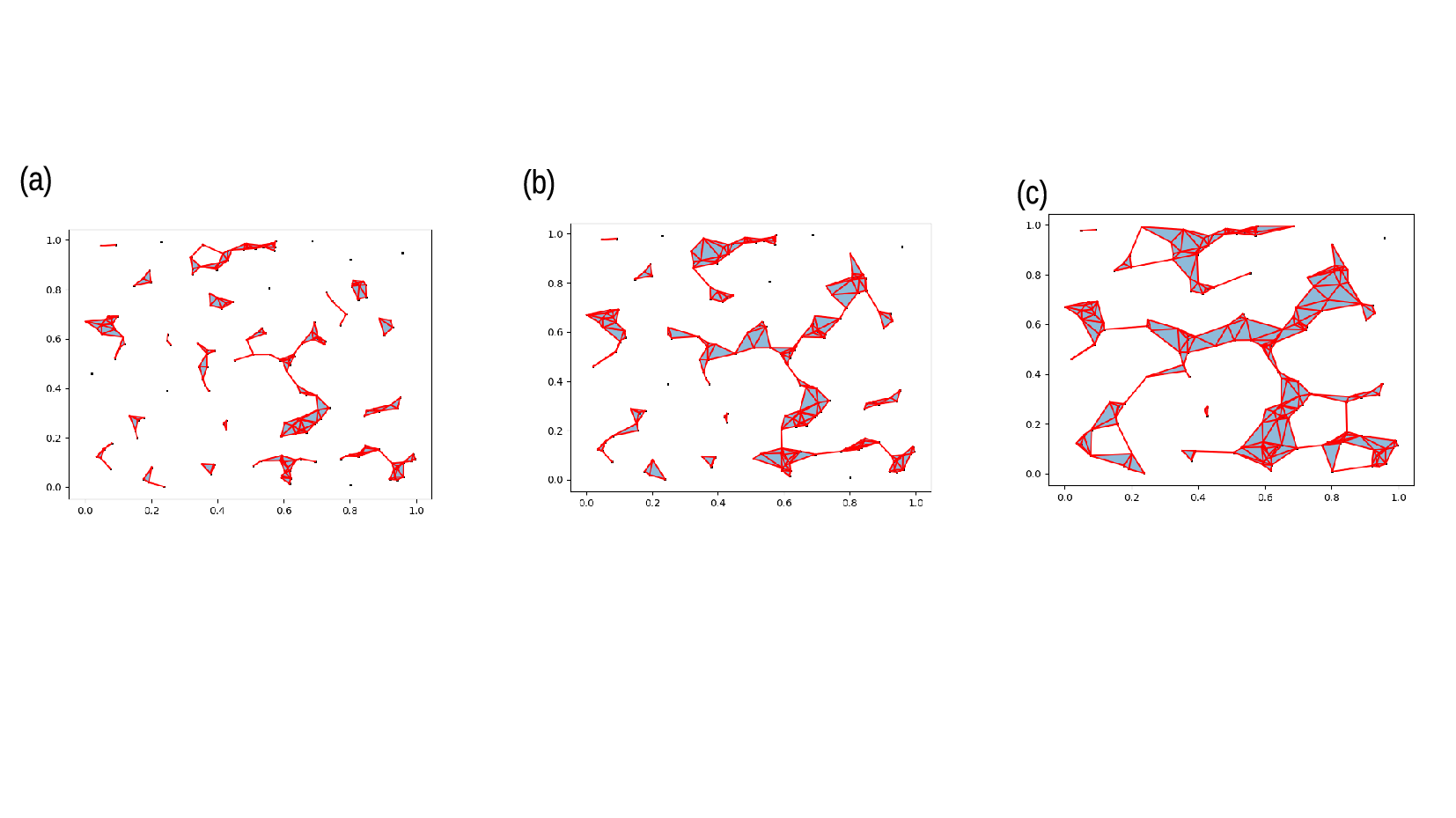}
\caption{ Alpha complexes of  point distribution at timestep t= 11 generated using Vicsek model[$ \eta=0, R=0.05$],(a) scale $r=0.03$,(b) $r= 0.045$, (c)$r=0.055$.}
\label{viscomplex_t11}
\end{figure}

We study two different situations of the dynamical point sets here-(i)the point sets with noise, $\eta = 0$ and (ii) the time-varying point sets with noise, $\eta=0.9$. We constructed the Alpha complexes with scale $r$ varying from $0.005 $ to $0.065$ unit. The scale of filtration to build these complexes is non-trivial as it highly influences the clusters and the loops. Fig.(\ref{viscomplex_t11}) shows the point cloud of the simulaton with noise , $\eta=0$ at time step $t=11$.The Alpha complex of the same point cloud at different scale $r$ generates different topological features as reflected in Fig. (\ref{viscomplex_t11}a-c). Hence, we consider finer variation of scale to construct the Euler Characteristic Surfaces, capturing sufficient topological information.

%{Verification of Proofs:}
We perform a numerical verification of the  the results  in Eq. \ref{SliceWiseStabilityEstimate-ECS-p=1} for these real simulated data-sets via construction of the Euler Characteristic Surfaces and comparing Euler $L_1$ metric with 1-Wasserstein distance. We estimated the metrics at  suitably spaced intervals of time-steps where the point sets have different geometry. 

Table.\ref{proof6_7} shows supporting evidence for the theory discussed in Section.\ref{PH}. Here, for a particular time step $t*$ and between two dynamical systems $K$ and $L$, 
let   $A= || \chi(\mathcal{K}(r, t*)) - \chi(\mathcal{L}(r, t*)) \|_1 $ and   $B= 2 \Big[\sum_{ 0 \le n \le 1} W_1(PD_{n} (\mathcal{K}),PD_{n} (\mathcal{L}))\Big]$ , where $ W_1$ is 1-Wasserstein distance between $n$ the persistence diagrams of $\mathcal{K}$ and $\mathcal{L}$.

According to Eq.(\ref{SliceWiseStabilityEstimate-ECS-p=1}), $A \le B$. We verified the relations between A and B  different time slices for Euler Characteristic Surfaces using two different dynamical systems (point cloud), one being the aggregation using Vicsek model for small noise $\eta=0$ and the other being the aggregation with comparatively large noise $\eta=0.9$. Some of the estimated values of $A$, $B$ are shown in Table.\ref{proof6_7}.

\begin{table}[h]

\caption{Measures between point clouds for $\eta=0.0 $ and $\eta=0.90$}\label{proof6_7}

\begin{tabular*}{\textwidth}{@{\extracolsep\fill}lcccccc}
\toprule% 
time$(t*)$ & point cloud at $\eta=0.0$ & point cloud at $\eta=0.9$ & A & B  \\
\midrule
 11 & \includegraphics[width=0.2\textwidth]{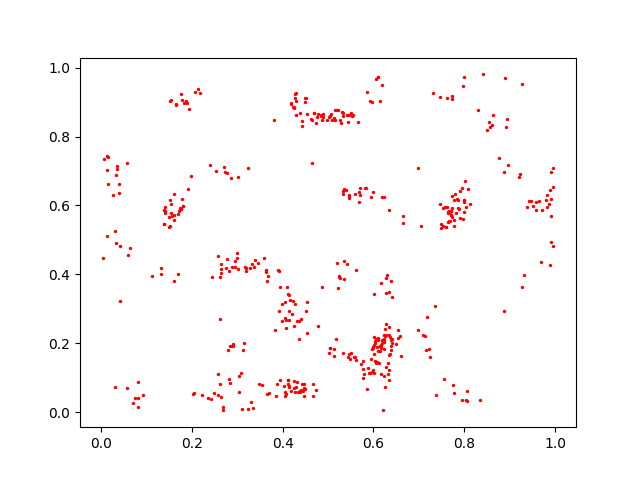} & \includegraphics[width=0.2\textwidth]{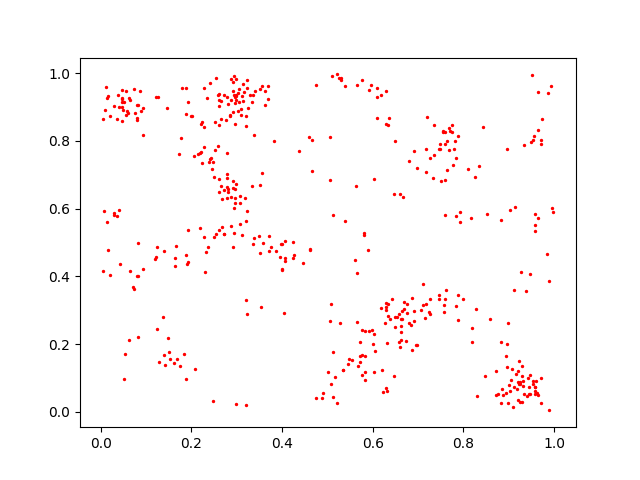}& 1.3869 & 2.3825 \\ 
 151 & \includegraphics[width=0.2\textwidth]{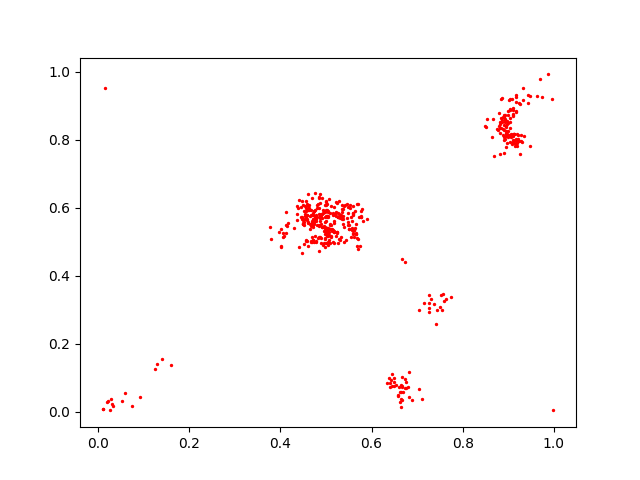} & \includegraphics[width=0.2\textwidth]{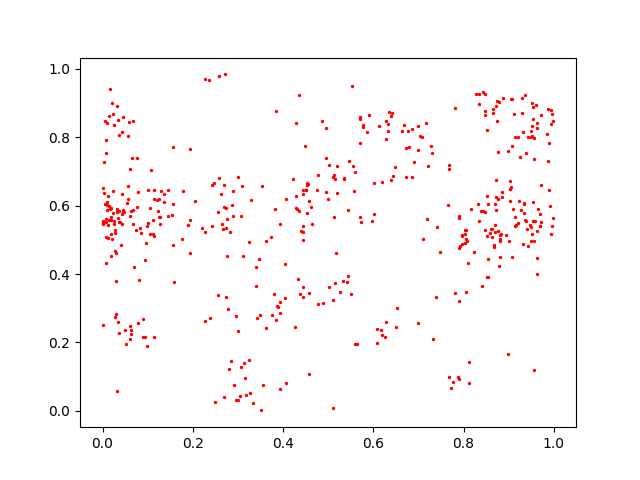}& 4.2975 & 6.3003 \\
 351 & \includegraphics[width=0.2\textwidth]{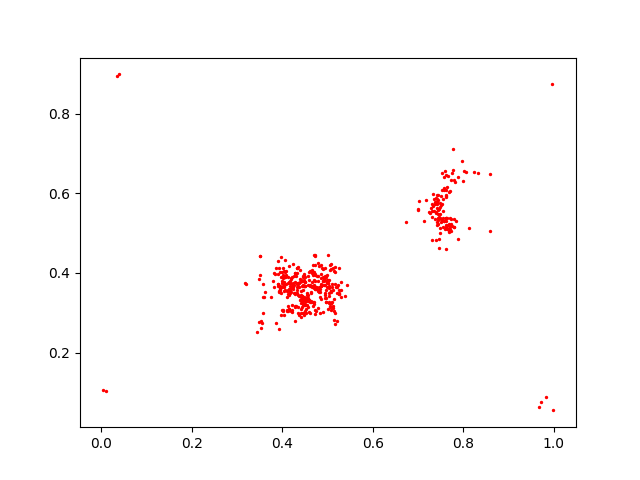} & \includegraphics[width=0.2\textwidth]{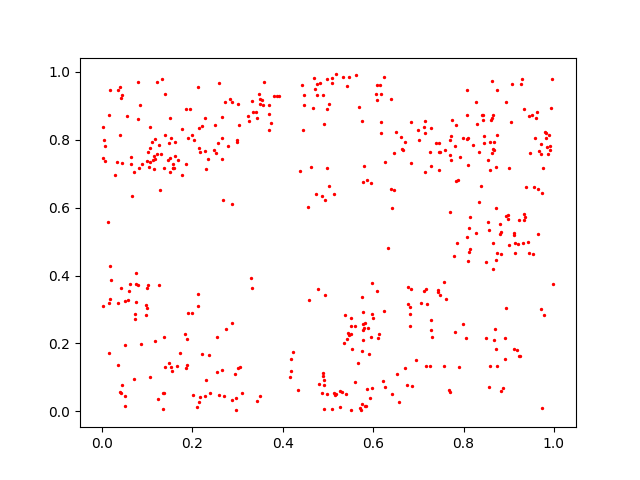}& 5.3040 & 7.5766 \\
 451 & \includegraphics[width=0.2\textwidth]{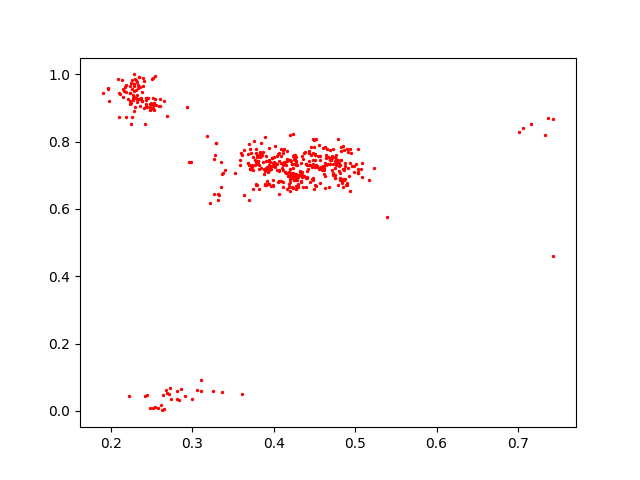} & \includegraphics[width=0.2\textwidth]{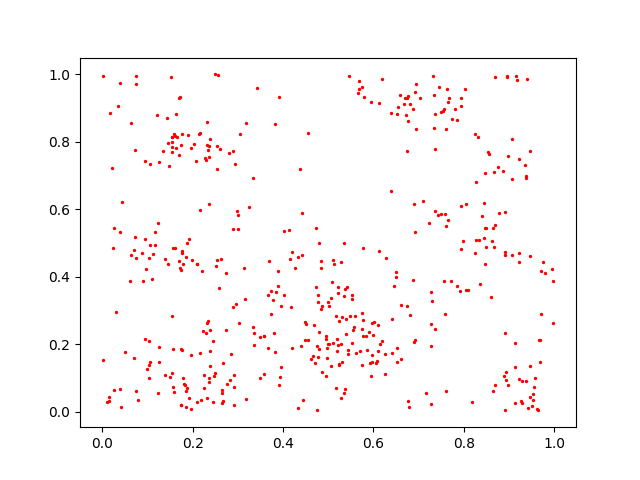}&5.1494 & 7.2694 \\  
\botrule 
\end{tabular*}

\end{table}
\noindent
\begin{figure}[h!]
\centering
\includegraphics[width=0.8\textwidth]{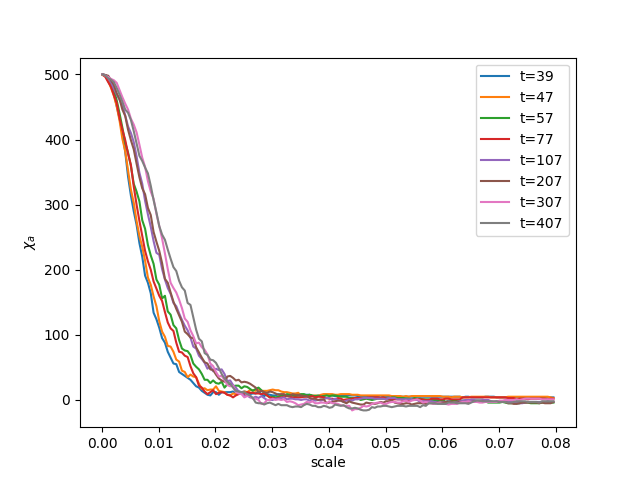}
\caption{Slices of Euler Characteristic Surfaces at different time steps for the ECS corresponding to $\eta=0.9$.}
\label{temp}
\end{figure}

\begin{figure}[h!]
\centering
\includegraphics[width=0.8\textwidth]{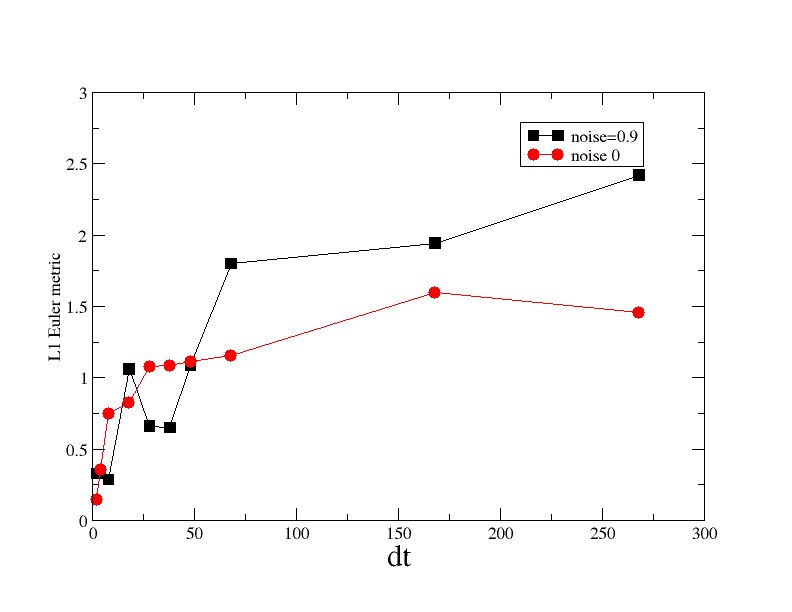}
\caption{The $L_{1}$ Euler metric between ECS time-slices with $dt$ distance apart from time slice $t=39$.}
\label{temp_em}
\end{figure}

We also analysed the temporal stability of our Euler Characteristic Surface with the simulated point sets. As the number of point is constant in every time-step of the simulation  and the point cloud moves with a finite and small velocity($\mid v \mid =0.03$), the change in Euler Characteristic Curves are hypothesized to be finite and small too. Fig.(\ref{temp}) shows the Euler Characteristic Curves, which are the time-slices of the Euler Characteristic Surface at different instants of time.

To quantify the temporal stability, the $L_{1}$ Euler metric between two Euler Characteristic Curves with different time-interval ($dt$) apart was calculated for the dynamic point clouds corresponding to $\eta=0.9$ and $\eta=0$. The smaller the time interval between two point clouds, the closer the Euler Characteritic curves are, Fig(\ref{temp}). This shows that the metric distance between two Euler Characteristic Curves, corresponding to two point clouds  that are small distances apart, is also small. This justifies the theorem about temporal stability of Euler Charactric Surface,theorem \ref{TemporalStabilityECS} and matches the quantitative bound. In Fig.(\ref{temp_em}), the L$_{1}$ Euler metrics between two Euler Characteristic Curves, $dt$ distance apart, are plotted against time interval $dt$. As one can see in the snapshots given in the Table (\ref{proof6_7}), when the simulation has no noise, $\eta=0$, the system gets quickly ordered forming dense and few clusters/flocks. For comparatively larger noise, $\eta=0.90$, the system does not follow such alignment, making the point cloud more scattered and sparse. This distinction of behaviour is reflected in the plot in Fig.(\ref{temp_em}). For zero noise the the slope of the  Euler $L_{1}$ metric is sharp and quickly reaches the equilibrium state. For noise $\eta=0.9$ the metric distance fluctuate initially and then takes longer time to reach a chaotic stable state. The chaos sustains without much fluctuation in the distribution of the point cloud.

\subsubsection{Analysing Point Clouds with different noises($\eta$), for different $R$ values:}
\begin{figure}[h!]
\centering
\includegraphics[width=0.99\textwidth]{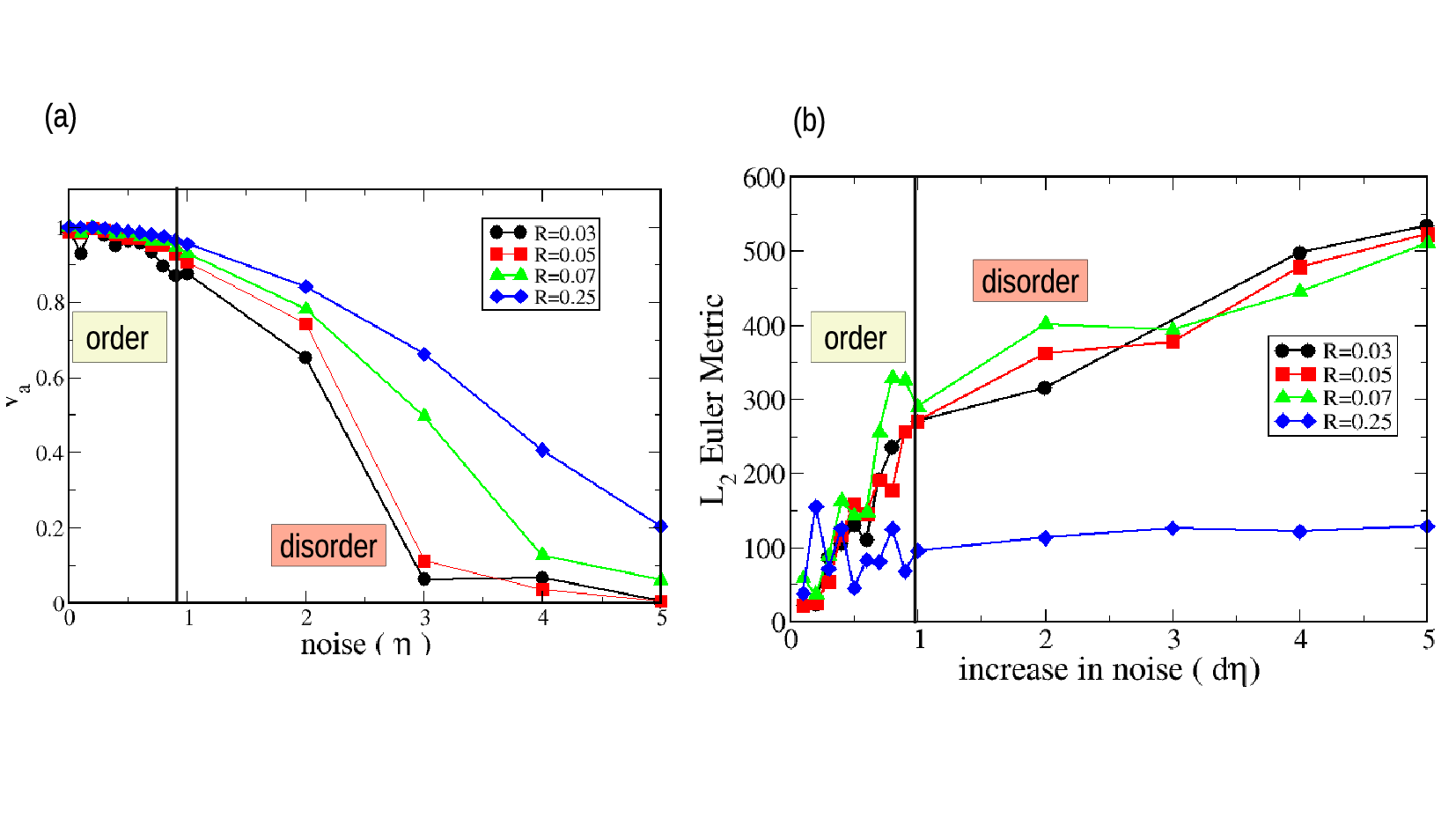}
\caption{ (a) $L_{2}$ Euler metric estimated between ECSs of noise $\eta$ and zero noise. (b) The order parameter vs  noise ($\eta$)plotted for different interaction radius $R$. }
\label{em_vis}
\end{figure}
Following this result from $L_{1}$ Euler Metric, we progressed with more analysis to understand the reflection of order-disorder transition on our topological marker Euler Metric between Euler Characteristic Surfaces. To define order in the system, we followed the traditional order parameter $v_{a}$ which is the average normalized velocity of the particles/points. Therefore $\displaystyle v_{a}= \frac{1}{N \mid v\mid}\mid \sum_{i=1}^{N} v_{i} \mid$. When the system is perfectly ordered or aligned in same direction, order parameter $v_{a} = 1$ and as the disorder increases it moves more towards zero. Fig.(\ref{em_vis}a) shows that with increase in noise $\eta$ how the system moves from order to disorder. We tried different variations in interaction radius $R$, measuring how far one particle/ point will be influenced by the orientation/movement of the neighbouring particles. For very large $R$ the system quickly reachs its equilibrium state (that may be ordered or disordered depending upon the value of noise) as the interaction is very high. Now, to look from the topological aspect, the Alpha complex of the dynamical point clouds corresponding to similar values of $\eta $ and $R$ was generated followed by the construction of Euler Characteristic Surfaces. We then compared the ECS corresponding to a specific $\eta$ with ECS corresponding to $\eta=0$, keeping the interaction radius $R$ constant. The $L_{2}$ Euler metric (Eq.\ref{Metric_eq}) was estimated for the comparison and the results are displayed in Fig.(\ref{em_vis}b). The plot shows that slope of Euler Metric remarkably changes around noise $\eta  \approx 0.9 $ where the order-disorder transition happens in correspondence to fig.(\ref{em_vis}a). This result suggests that once the system enters into disorder with increase in noise, the metric distance vary less, making a plateau like region in the plot of Euler metric.  Thus, our topological tools, Euler Characteristic Surfaces and Euler metric, are not merely a measure that only estimates similarity and dissimilarity between two dynamical systems but also signals the physical phase transition. This need to be studied more in details in future to explore whether quantitative changes in ECS measured by the Euler Metric can be introduced as an alternative marker to characterize phase transition in collective motion phenomena.      

\section{Conclusion}\label{conc}
This work establishes the robustness of our proposed construction of  the spatiotemporal topological map of a dynamical system - the Euler Characteristic
Surface (ECS), and a measure that can differentiate between two ECSs - the Euler $L_1$ Metric.  We show the correlation between the tools of ECS construction with the standard measure of Persistent Homology (PH)  commonly used by the TDA community. Classically, PH examines the dynamical system at different length scales at any fixed time, while most dynamical systems show interesting topological changes in both scale and time. In recent times, the time development of a dynamical system has been analyzed using Crocker Plots, Persistent Vineyards, etc, bringing Persistent Homology into dynamic situations. These methods are computationally expensive - as persistent homology is -  and their stability needs to be explored further. Our proposed topological tool of the ECS  is an attempt to encapsulate information of both scale and time of a dynamical system in a single map. On establishing the stability and robustness to perturbation for ECS here, the construction does emerge as powerful topological tool for analyzing dynamical systems.

The  highlights of  this work may be summarized thus:
\begin{itemize}
\item Estimating ECS using appropriate simplicial complexes based on the  data points (filtered by a scale parameter) gives more precise results than by using discretized grids to map data cloud and calculate Euler characteristic, as has been often done by the image processing community. However in the case of extracting data from digital images where estimation of Euler characteristic often involves usage of discrete grids, the ECS from both approaches were close. Preferably, ECS construction using simplicial complexes appears the best choice for large scale systematic analysis.

\item The Euler Characteristic Surfaces are robust to small perturbations in scale and time provided they are built up with fine resolutions in scale and time and up to sufficiently large scales that cover almost all topological features (all critical scales) and changes present in the dynamical system. 

\item We recreated the Euler Characteristic Surfaces via Betti Curves from Persistence Diagrams and established the correlation between a time-slice of the spatiotemporal ECS and Persistence Diagrams. This makes the ECS an alternative and more powerful tool of analysis as it also contains the temporal development of the system.

\item The relations between the metric distances, the p-Wasserstein distance of PH  and the $L_1 $ and $L_2$ Euler Metric of ECS, lends to the stability of the ECS. The Euler $L_2$  Metric has a Hilbert space structure making it  suitable for direct applications in machine learning algorithms, while better stability results are obtained using the Euler $L_1 $ Metric. However, for systems with finite number of points this is not a major disadvantage as $L_ 1$  is embeddable in Hilbert space with bounded distortion.

\item The ECS with the $L_1$ metric is demonstrated to be an efficient summary to quantify the similarities and dissimilarities between two different data sets. 

\item The ECS construction with the $L_1$ metric can give cues to order-disorder transitions in the collective motion of particles as was observed in an application of particle dynamics using the Vicsek Model. This was also observed in our application of particle dynamics using a modified egg-beater flow where critical points of topological changes could be captured using this metric measure.

\end{itemize} 

The advantage of  the ECS is its ability to give a topological summary of big and complex time varying data sets at a minimal computational expense and a quantifiable stability. The successful application of the tools and the output measures on continuous and finite time varying metric spaces of both the Vicsek model and the eggbeater flow model establishes the robustness of our constructions.

The construction and application of Euler Characteristic Surface for time series data is relatively recent and  therefore requires more theoretical studies and more applications on different dynamics for its successful interpretation. However the stability guaranties that emerge from its mathematical framework and  its ability to signal phase transitions in dynamical processes is suggestive of its potential to be used as an efficient spatiotemporal topological summary. 

While theorem \ref{SliceWiseStability-ECS} and  theorem \ref{TemporalStabilityECS} are stability results  for Euler Characteristic Surfaces for scale and time variations, a natural generalization of the above stability results is an investigation of the combined stability of a possibly modified Euler Characteristic Surface over a finite scale and time domain $\Omega = [0, R] \times [0,T]$.  A pertinent project is an investigation of stability results of a modified Euler Characteristic Surface with appropriate metric with respect to  various constructions that have been proposed for persistent homology of time-varying spaces (as  mentioned in the introduction)  such as  crocker plots, persistent vineyards, crocker stacks, zigzag persistence.  We are exploring those considerations as a followup project.
\bibliography{ph_arxiv}
\end{document}